\documentclass[aps,prd,preprint]{revtex4}
\usepackage{amsfonts}
\usepackage{amssymb}
\usepackage{amsmath}
\usepackage{graphicx}
\usepackage{graphics}
\usepackage{color}
\usepackage{epsfig}
\newcommand{\be}{\begin{equation}}
\newcommand{\ee}{\end{equation}}
\newcommand{\bea}{\begin{eqnarray}}
\newcommand{\eea}{\end{eqnarray}}
\newcommand{\bwt}{\begin{widetext}}
\newcommand{\ewt}{\end{widetext}}
\begin{document}
%%%%%%%%%%%%%%%%%
%%%%%%%%%%%%%%%%%

\title{Holographic Walking Technicolor and Stability of Techni-Branes}
\author{T.E. Clark}
\email[e-mail address:]{clarkt@purdue.edu}
\affiliation{Department of Physics,\\
 Purdue University,\\
 West Lafayette, IN 47907-2036, U.S.A.}
\author{S.T. Love}
\email[e-mail address:]{loves@purdue.edu}
\affiliation{Department of Physics,\\
 Purdue University,\\
 West Lafayette, IN 47907-2036, U.S.A.}
\author{T. ter Veldhuis}
\email[e-mail address:]{terveldhuis@macalester.edu}
\affiliation{Department of Physics \& Astronomy,\\
 Macalester College,\\
 Saint Paul, MN 55105-1899, U.S.A.}

\vspace*{1.0in.}

\begin{abstract}
Techni-fermions are added as stacks of $D7$-$\overline{D7}$ techni-branes within the framework of a holographic technicolor model that has been proposed as a realization of walking technicolor.  The stability of the embedding of these branes is determined.  When a sufficiently low bulk cut-off is provided the fluctuations remain small.  For a longer walking region, as would be required in any realistic model of electroweak symmetry breaking, a larger  bulk cut-off is needed and in this case the oscillations destabilize.
\end{abstract}

\maketitle

\section{Introduction}

\begin{figure*}
\begin{center}
$\begin{array}{cc}
\includegraphics[scale=1.0]{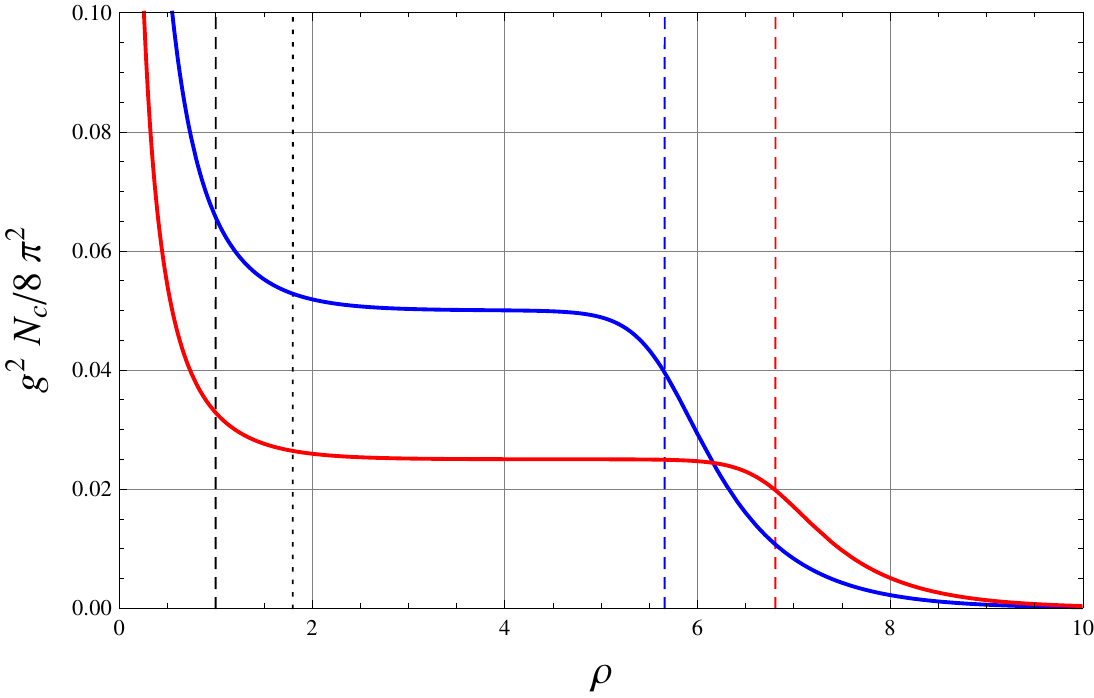}
\end{array}$
\caption{The running 't Hooft gauge coupling constant $g^2 N_C/8\pi^2$ as a function of the logarithm of the renormalization scale which is related to the distance $\rho$ into the bulk.  Each plot is for $N_C =10$ and the upper blue curve corresponds to $C=50$, $T=10^{-10}$ while the lower red curve corresponds to $C=100$, $T=10^{-12}$.  The smaller the value of $C$, the stronger the gauge coupling in the walking region.  The smaller the value of $T$, the greater the length of the walking region.  As seen in the plot, the walking region extends from $\rho \approx 1$, as indicated by the vertical black dashed line, to, for $T=10^{-10}$, $\rho\approx\rho_* = 5.66$, as indicated by the vertical blue dashed line, and, for $T=10^{-12}$, $\rho\approx\rho_* = 6.81$ as indicated by the vertical red dashed line.  The black dotted vertical line indicates a typical initial walking region location corresponding to the $D7$-$\overline{D7}$ overlap at $\rho_0 =1.8$.}
\label{fig:RunningCoupling}
\end{center}
\end{figure*}

Walking technicolor, a non-Abelian gauge theory exhibiting nearly-conformal behavior over a certain finite energy range, offers the potential to provide a phenomenologically consistent description of dynamical electroweak symmetry breaking \cite{Holdom:1984sk,Hill:2002ap}.  This scheme has been argued to be consistent with the range of values of the electroweak precision parameters while allowing for an extended Yukawa sector with acceptable flavor changing neutral currents \cite{Lawrance:2012cg,Foadi:2012ga}.  Characteristic of such models are distinct regions of renormalization group behavior.  At high energy, the gauge coupling constant exhibits asymptotic freedom.  As the distance scale increases the coupling enters an approximate scale invariance region of strong coupling where it slows down (\lq\lq walking" region).  At still larger distances, the gauge coupling begins to grow rapidly until it enters the confinement region where there is no longer any vestige of the scale symmetry.  Such renormalization group walking behavior was first demonstrated using the perturbative 2-loop Banks-Zaks $\beta$-function in which a large number of fermions provided the necessary cancellation to produce a small $\beta$ \cite{Banks:1981nn}.  In addition to direct field theory searches for walking $\beta$-functions, the gauge/gravity duality provides an alternative means to calculate the running in non-perturbative regions of coupling constant space.  Indeed, Nunez, Papadimitriou and Piai \cite{Nunez:2008wi} constructed a Type IIB supergravity background dual to a strongly coupled ${\cal{N}}=1$ SUSY gauge theory which exhibits the above described running behavior.  This running occurs without introducing techni-flavors and as such is a non-Banks-Zaks renormalization group behavior.  The background is obtained by considering a supergravity limit of a stack of $N_C$ $D5$ branes wrapping a 2-cycle resulting in a 10 dimensional space-time with the interval %to leading order in an expansion in terms of $N_c /C$%
\bea
ds^2 &=& \left( \frac{1}{64 C^3 T}\right)^{1/4} \left[ dx^2_{1,3} + \frac{P^\prime (\rho)}{8}\left(4 d\rho^2 +(\cos{\theta}d\varphi +\tilde\omega_3)^2 \right)\right.\cr
 & &\left. +\frac{P(\rho) \coth{2\rho}}{4}\left(d\theta^2 +\sin{\theta}^2 d\varphi^2 +\tilde\omega_1^2 +\tilde\omega_2^2 \right.\right. \cr
 & &\left.\left.\qquad\qquad\qquad\qquad +\frac{2}{\cosh{2\rho}}\left( \tilde\omega_1 d\theta -\tilde\omega_2 \sin{\theta} d\varphi \right)  \right) \right].
\label{10dinterval}
\eea
Here $C$ and $T$ are integration constant parameters and the $SU(2)$ left-invariant coordinates for the $S^3$ are given by
\bea
\tilde\omega_1 &=& \cos{\psi} d\tilde\theta + \sin{\psi} \sin{\tilde\theta} d\tilde\varphi \cr
\tilde\omega_2 &=& -\sin{\psi} d\tilde\theta + \cos{\psi} \sin{\tilde\theta} d\tilde\varphi \cr
\tilde\omega_3 &=& d\psi + \cos{\tilde\theta} d\tilde\varphi .
\eea
The function $P(\rho)$ (with $P^\prime (\rho) =\frac{d}{d\rho}P(\rho)$) obeys a BPS obtained master differential equation which in this supergravity limit has the leading order small $N_C/C$ solution
\be
P(\rho)=4C\left[1 + 3T\left(\sinh{4\rho} -4\rho  \right)  \right]^{1/3} .
\ee
%with parameters $C=c \cos{\alpha}/4$ and $T=\tan^3{\alpha}/3$ constrained to be $1<<\cot{\alpha}\leq e^{(2^{4/3}/3)  (c/N_c)}$ in this approximation.%
Using a $D5$ probe brane to define an appropriate 4-dimensional 't Hooft gauge coupling constant  \cite{Nunez:2008wi},
\be
\frac{g^2 N_C}{8\pi^2}=\frac{N_C \left(  \cosh{2\rho} +1\right)}{4C \sinh{2\rho}}\frac{1}{[1+3T(\sinh{4\rho} -4\rho)]^{1/3}},
\label{gaugecoupling}
\ee
whose renormalization group behavior  (in $\rho$) was shown to exhibit a walking region for $1\leq \rho \leq \rho_* \approx \frac{1}{4}\ln{(2/3T)}>>1$ with confinement for small $\rho < 1$ and asymptotic freedom for $\rho$ in the UV, $\rho >\rho_*$.  The walking region is seen to begin for $\rho$ values for which the $\coth{2\rho}$ prefactor in equation (\ref{gaugecoupling}) starts to become constant.  This is approximately for $\rho=1$.  The gauge coupling remains nearly constant until $3 T \sinh{4\rho}$ begins to grow to become comparable to 1.  Then the coupling constant
begins to decrease in the field theory UV region.  Thus the walking region ends approximately at 
$3Te^{4\rho_*}/2 \approx 1$.  Since the value of $T$ controls the size of the walking region its value is chosen to allow there to be a significant energy difference between the confinement region and the UV asymptotic freedom region.  The renormalization group running of the gauge coupling is depicted in Fig.( \ref{fig:RunningCoupling}).

Anguelova, et al. \cite{Anguelova:2011bc} added techni-flavors to the model by embedding $N_F$ stacks of $D7$ and $\overline{D7}$ branes in a $U$-shaped geometry necessary for $U(N_F) \times U(N_F) \rightarrow U(N_F)$ symmetry breakdown \cite{Sakai:2004cn}.  They employed holographic techniques in \cite{Anguelova:2011bc} in order to study the vector and axial vector meson contributions to the electroweak precision parameters.  

The purpose of this paper is to analyze the stability of such a $D7$-$\overline{D7}$ techni-brane embedding \cite{Anguelova:2011bc,Anguelova:2012ka}.   Reference \cite{Anguelova:2011bc} showed that the squared masses of the vector mesons were all positive, while in reference \cite{Anguelova:2012ka}, the squared masses of the scalar and pseudoscalar mesons were evaluated and again claimed to all be positive. In the present paper, we critically reanalyze the scalar and pseudoscalar mass spectra. While we concur that the pseudoscalar squared masses are positive, we find that, except for cases where the UV cut-off is such as to allow only a very small walking region, the lowest scalar mass squared is negative thus indicating that the brane embedding is unstable. 

\section{Techni-brane embedding}

\begin{figure*}
\begin{center}
$\begin{array}{cc}
\includegraphics[scale=0.80]{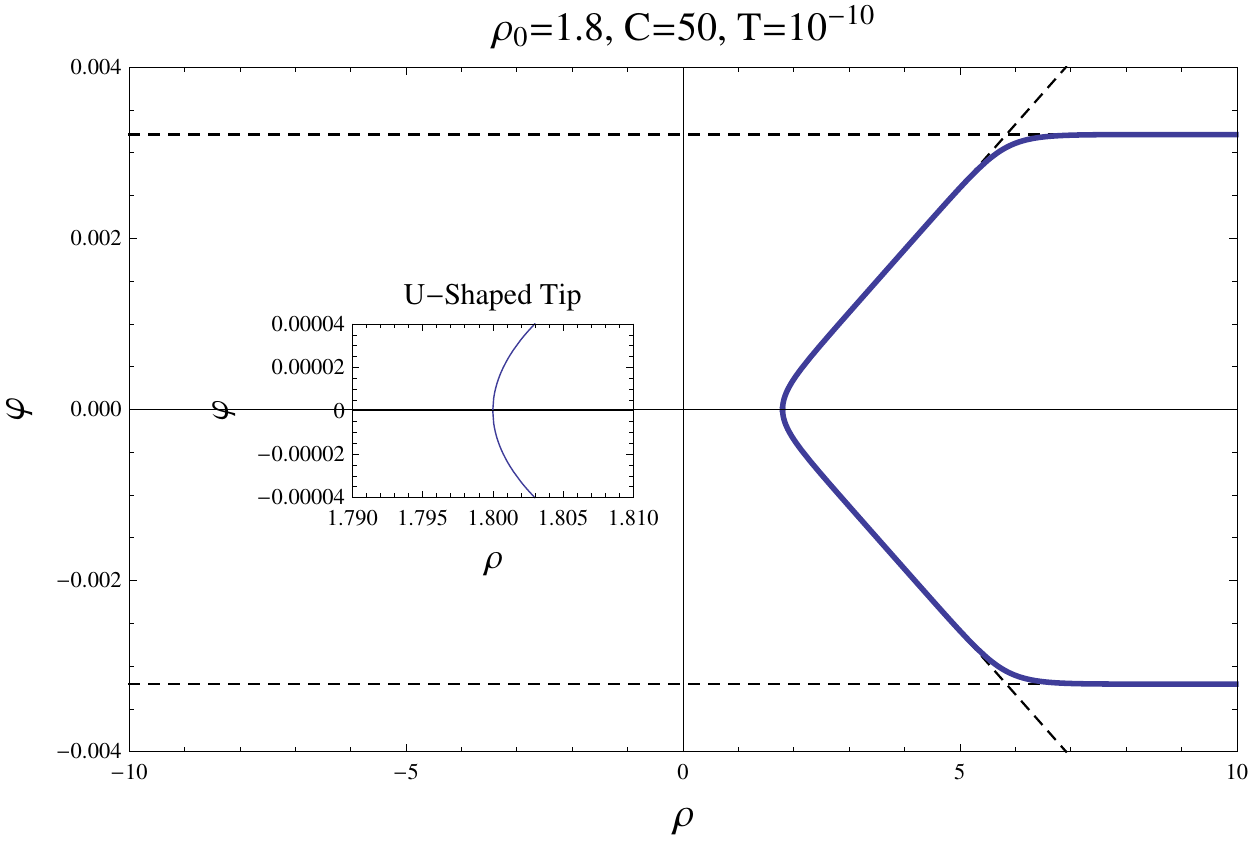} \\
\includegraphics[scale=0.75]{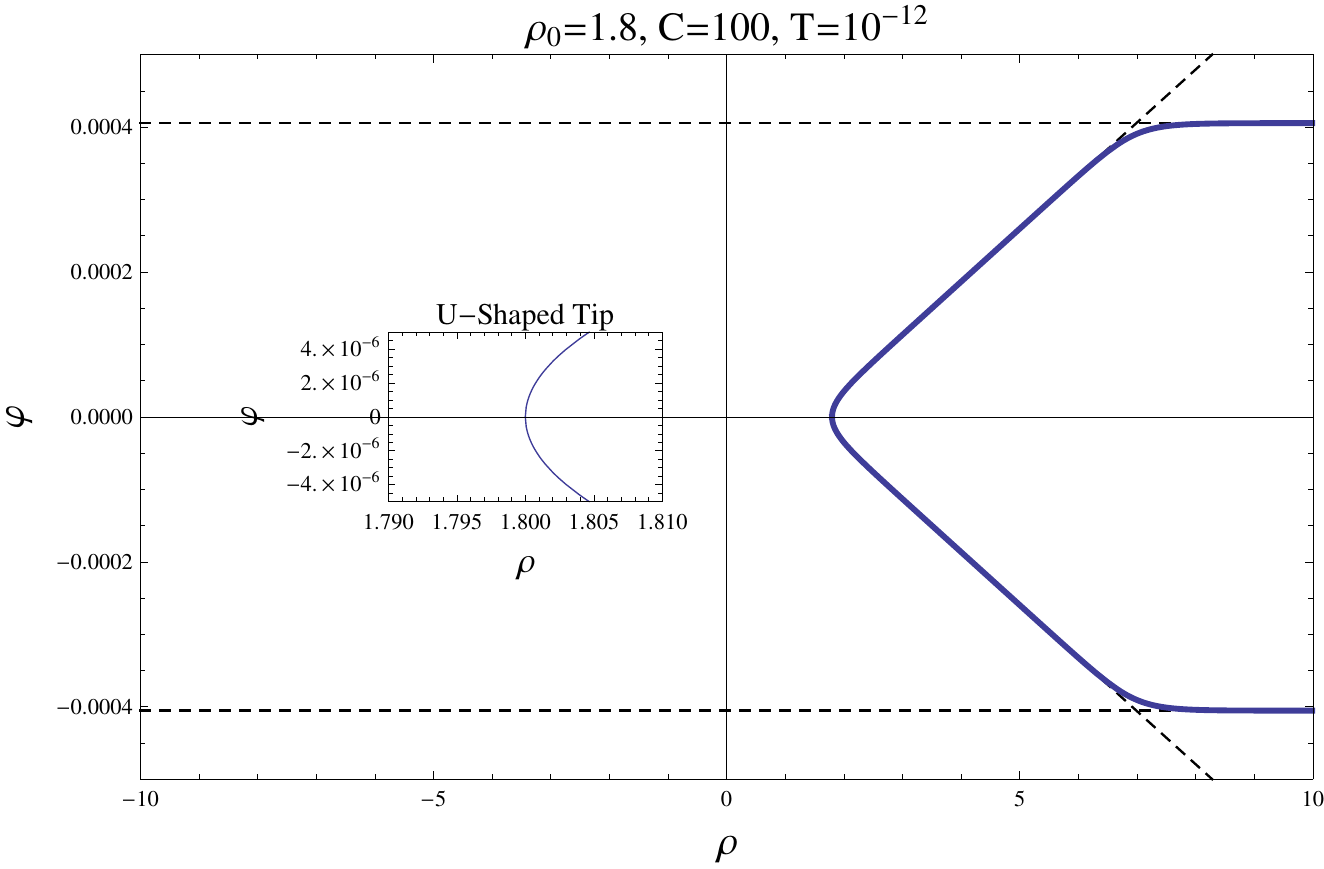}
\end{array}$
\caption{Embedding functions for the $D7$-$\overline{D7}$ branes.  The $D7$ branes are embedded along the positive $\varphi$ curves and the $\overline{D7}$ branes are embedded along the negative $\varphi$ curves.  The branes smoothly overlap at the U-shaped tip of the embedding curves at $\rho = \rho_0 =1.8$ as more closely depicted in the magnified inserted plots.  The numerical values of the asymptotic embedding angles in the upper plot are $\varphi_\pm =\pm 0.00321105$ and in the lower plot are $\varphi_\pm =\pm 0.000405376$.  The dashed oblique lines are the explicit embedding profiles given by equation (\ref{profile}) obtained from the simplified metric in the walking region.}
\label{fig:D7-embedding}
\end{center}
\end{figure*}

A $D7$ probe brane is embedded as in \cite{Anguelova:2011bc, Anguelova:2012ka} according to which the brane profile is obtained by solving the field equations obtained from the Dirac-Born-Infeld (DBI) action of the $D7$ brane with induced metric $g_{(8)}$.  The branes are embedded according to Table (1) with the $D7$ brane having the complementary coordinates as functions of the $\rho$ coordinate, $\theta = \theta (\rho)$ and $\varphi =\varphi (\rho)$.  
\begin{center}
\begin{tabular}{||c||c|c|c|c|c|c|c|c|c|c|c|}
\hline
$ x^M$, $M=0 ,\dots, 9$ & $x^0$~ & ~$x^1$~ & ~$x^2$~ & ~$x^3$~ & ~$\theta$~ & ~$\varphi$~ & ~$\rho$~ & ~$\tilde{\theta}$~ & ~$\tilde{\varphi}$~ & ~$\psi$~ \cr
\hline
$D5$&x&x&x&x&x&x& & & & \cr
\hline
~$D7$-$\overline{D7}$~~&x&x&x&x& & &x&x&x&x\cr
\hline
\end{tabular}\\
\vspace*{6pt}
Table 1.
\end{center}
The associated DBI Lagrangian is \cite{Anguelova:2011bc} 
\be
{\cal L}= L_0 \sqrt{-\det{g_{(8)}}}=L_0\left(f(\rho) + g(\rho) \left[ \theta_\rho^2 (\rho) + \sin^2{\theta (\rho)} \varphi_\rho^2 (\rho)\right]  \right)^{1/2} ,
\ee
with $L_0$ a constant and the subscript $\rho$ denoting differentiation $\frac{d}{d\rho}$ while 
\bea
f(\rho) &=& \frac{1}{16^5 C^6 T^2} P^2 (\rho) P_\rho^2 (\rho) \coth^2{2\rho} = \frac{1}{16^2 C^2} \frac{\sinh^2 {4\rho}}{[1+3T(\sinh{4\rho} -4\rho)]^{2/3}}\cr
g(\rho) &=& \frac{f(\rho) P(\rho)}{2 P_\rho (\rho) \coth{2\rho}}=\frac{2 \sinh{4\rho} [1+3T(\sinh{4\rho} -4\rho)]^{1/3}}{16^3 C^2 T}.
\label{fullfandg}
\eea
Note that $\varphi$ is a cyclic coordinate, $\frac{\partial {\cal L}}{\partial \varphi}=0$, so the field equations lead to the embedding equations for the $D7$ brane
\bea
\theta &=& \pi/2 \cr
 & & \cr
\frac{\partial {\cal L}}{\partial \varphi_\rho} =\frac{g(\rho)\varphi_\rho}{\sqrt{f(\rho) +g(\rho)\varphi_\rho^2}} &=& \sqrt{g(\rho =\rho_0)}\equiv \sqrt{g_0} ,
\label{embedding1}
\eea
where the integration constant is defined by the parameter $\rho_0$ with $g(\rho_0)=\frac{2 \sinh{4\rho_0} [1+3T(\sinh{4\rho_0} -4\rho_0)]^{1/3}}{16^3 C^2 T}\equiv g_0$, where in contrast to \cite{Anguelova:2011bc} no additional relation among the parameters $\rho_0$, $C$, and $T$ is needed.  The $D7$ and $\overline{D7}$ branes overlap at $\rho=\rho_0$ and their location on this $U$-shaped embedding equation is for $\rho > \rho_0$ obtained from equation (\ref{embedding1}),
\be
\frac{d \varphi (\rho)}{d\rho}=\left[ \left(\frac{f(\rho)}{g(\rho)}\right) \frac{g_0}{g(\rho) -g_0} \right]^{1/2}  ,
\ee
and integrating once more to find $\varphi =\varphi (\rho)$
\be
\varphi (\rho)= \pm \int_{\rho_0}^\rho d\rho^\prime \sqrt{\left(\frac{f(\rho^\prime)}{g(\rho^\prime)}\right) \left[\frac{g_0}{g(\rho^\prime) -g_0} \right]}  .
\label{embedding2}
\ee
The embedding solution is depicted in Fig.( \ref{fig:D7-embedding}) where it is seen that the azimuthal angle $\varphi (\rho)$ approaches asymptotic values $\varphi_\pm = \pm |\varphi (\infty)|$ as $\rho$ goes beyond the walking region $\rho > \rho_*$, while it smoothly approaches its turning point at $\rho=\rho_0$.

For values of $\rho_0$ and $\rho$ in the walking region, the embedding profile can be found explicitly.  The functions $f(\rho)$ and $g(\rho)$ simplify to become
\be
f(\rho) \approx \frac{e^{8\rho}}{4^5 C^2} \qquad, \qquad
g(\rho) \approx \frac{e^{4\rho}}{4^6 C^2 T},
\label{walkingfandg}
\ee
with $g_0 =g(\rho_0)\approx e^{4\rho_0}/4^6 C^2 T$, yielding a simple integral for $\varphi (\rho)$
\bea
\varphi (\rho)&=&\pm \int_{\rho_0}^{\rho}d\rho^\prime \frac{\sqrt{4T} e^{2\rho^\prime}}{\sqrt{e^{4(\rho^\prime -\rho_0)}-1}} =\pm \sqrt{T} e^{2\rho_0} \int_1^{e^{2(\rho -\rho_0)}}\frac{dx}{\sqrt{x^2 -1}}\cr
 & & \cr
 &=&\pm \sqrt{T} e^{2\rho_0}\cosh^{-1}{e^{2(\rho -\rho_0)}}.
\eea
Hence, the embedding profile can be expressed as \cite{Anguelova:2011bc}
\be
\tanh{\left[ \frac{\varphi e^{-2\rho_0}}{\sqrt{T}} \right]} = \pm \sqrt{1-e^{-4(\rho-\rho_0)}} .
\label{profile}
\ee
The asymptotic values for the embedding angles $\varphi_\pm$ can be estimated by evaluating the profile at the end of the walking region
\be
\varphi_\pm \approx \pm \varphi (\rho_*)= \pm \sqrt{T} e^{2\rho_0} \left[\ln{2} + 2(\rho_* -\rho_0)  \right].
\ee

For $\rho >\rho_{**}>\rho_*$ in the UV region where typically $\rho_{**}-\rho_* = 0.4$ will be taken, the metric also simplifies so that the functions $f(\rho)$ and $g(\rho)$ become
\be
f(\rho) \approx \frac{(\frac{2}{3T})^{2/3}}{4^5 C^2} e^{16\rho /3}\qquad, \qquad
g(\rho) \approx \frac{(\frac{2}{3T})^{-1/3}}{4^6 C^2 T}e^{16\rho /3}  .
\ee
With $\rho_0$ in the walking region so that $g_0 \approx\frac{e^{4\rho_0}}{4^6 C^2 T}$, the embedding profile for large $\rho$, with $\rho_{*}<\rho_{**}<\rho$, yields the approximate relation
\be
\varphi (\rho)= \varphi (\rho_{**})+\sqrt{\frac{3}{8}} e^{\frac{2}{3}\rho_*} e^{2\rho_0}\left[e^{-\frac{8}{3}\rho_{**}} - e^{-\frac{8}{3}\rho}\right].
\ee
A refined approximation for the asymptotic angles $\varphi_\pm$ can be obtained by expressing $\varphi (\infty)$ as
\be
\varphi (\infty)= \varphi (\rho_{**})+ \varphi_{\rm UV}  =\varphi (\rho_{**})+ \sqrt{\frac{3}{8}} e^{-\frac{8}{3}(\rho_{**}-\rho_*)} e^{2(\rho_*-\rho_0)}.
\label{approxasymp}
\ee
Further approximating $\varphi(\rho_{**})$ using the walking region profile and a small step correction
\be
\varphi(\rho_{**})= \varphi(\rho_*) + \Delta\varphi ,
\ee
where equation (\ref{embedding2}) can be used to yield
\bea
\Delta\varphi &=& \int_{\rho_*}^{\rho_{**}} d\rho^\prime \sqrt{\left(\frac{f(\rho^\prime)}{g(\rho^\prime)}\right) \left[\frac{g_0}{g(\rho^\prime) -g_0} \right]}  \approx \left( \rho_{**} -\rho_* \right)\sqrt{\left(\frac{f(\rho_{**})}{g(\rho_{**})}\right) \left[\frac{g_0}{g(\rho_{**}) -g_0} \right]} \cr
 &\approx& \left( \rho_{**} -\rho_* \right) \sqrt{\frac{8}{3}} e^{-\frac{8}{3}(\rho_{**}-\rho_*)} e^{2(\rho_*-\rho_0)} .
\eea
The various contributions to the asymptotic angles for the cases depicted in Fig.( \ref{fig:D7-embedding}) are listed in Table (2) in which $\rho_{**} =\rho_* +0.40$ is used and the numerical integration of equation 
(\ref{embedding2}) for $\varphi (\infty)$ is presented as the final column entry.
\begin{center}
\begin{tabular}{||c|c||c|c|c|c|c|c|}
\hline
 $T$&~$\rho_*$~ & ~$\varphi(\rho_*)$~ & ~$\Delta\varphi$~ & ~$\varphi(\rho_{**})$~ & ~$\varphi_{\rm UV}$~ & ~$\varphi (\infty)$~ & ~Method~ \cr
\hline
\hline
$10^{-10}$&5.66&0.0030791&0.0000998&0.0031788&0.0000935&0.0032723&Approximate~\cr 
\hline
$10^{-10}$&5.66&0.0029884&0.0001339&0.0031223&0.0000888&0.0032111&Numerical~\cr
\hline
\hline
$10^{-12}$&6.81&0.00039208&0.00001000&0.00040208&0.00000938&0.00041146&Approximate~\cr
\hline
$10^{-12}$&6.81&0.00038305&0.00001342&0.00039647&0.00000891&0.00040538&Numerical~\cr
\hline
\end{tabular}\\
\vspace*{6pt}
Table 2.
\end{center}

\section{Techni-Brane Instability}

In order to determine the stability of this probe brane embedding, small oscillations about the $D7$ brane profile are considered in the $(x^\mu , \rho)$ directions \cite{Kuperstein:2008cq}, \cite{Kruczenski:2003uq}
\bea
\theta (x, \rho) &=& \pi/2 + \Delta\theta (x, \rho) \cr
\varphi (x, \rho) &=& \varphi (\rho) + \Delta\varphi (x, \rho) ,
\eea
where $\varphi (\rho)$ is the above embedding solution given by equation (\ref{embedding2}).  Substituting these along with their differentials, $d\theta (x,\rho) = d\Delta\theta=d x^\mu \partial_\mu \Delta\theta + d\rho \frac{\partial}{\partial \rho} \Delta\theta$ and $d\varphi (x, \rho)= d\varphi (\rho) +d\Delta\varphi (x, \rho)= d\rho \frac{\partial}{\partial \rho} \varphi (\rho) + d x^\mu \partial_\mu \Delta\varphi + d\rho \frac{\partial}{\partial \rho} \Delta\varphi$, into the equation (\ref{10dinterval}), the induced $D7$ brane metric and hence DBI action are obtained.  Expanding the DBI action for the $D7$ brane to quadratic order in the fluctuations $\Delta\theta$ and $\Delta\varphi$ yields the fluctuation Lagrangian
\bea
{\cal L}&=& -\frac{1}{2} a_\varphi (\rho) \partial_\mu \Delta\varphi \partial^\mu \Delta\varphi -\frac{1}{2} b_\varphi (\rho) \left( \frac{\partial}{\partial \rho} \Delta\varphi \right)^2 -\frac{1}{2}v_\varphi (\rho) \Delta\varphi^2 \cr
 & &-\frac{1}{2} a_\theta (\rho) \partial_\mu \Delta\theta \partial^\mu \Delta\theta -\frac{1}{2} b_\theta (\rho) \left( \frac{\partial}{\partial \rho} \Delta\theta \right)^2 -\frac{1}{2}v_\theta (\rho) \Delta\theta^2 ,
\eea
where the Lagrangian coefficients are given by
\bea
a_\varphi (\rho) &=& \frac{f(\rho) P(\rho)}{4\coth{2\rho} \sqrt{f(\rho) +g(\rho) \varphi^2_\rho}} =\frac{16^3 C^3 T}{4\cosh^2{2\rho}} \sqrt{f(\rho) g(\rho)}\sqrt{ [g(\rho)-g_0]}  \cr
b_\varphi (\rho) &=& \frac{f(\rho) g(\rho)}{[f(\rho) +g(\rho) \varphi^2_\rho]^{3/2}}=\frac{[g(\rho)-g_0]^{3/2}}{\sqrt{f(\rho) g(\rho)}}   \cr
a_\theta (\rho) &=& \frac{P(\rho) \sqrt{f(\rho) +g(\rho) \varphi^2_\rho}}{4\coth{2\rho}}=\frac{g(\rho)}{[g(\rho)-g_0]}~  a_\varphi   \cr
b_\theta (\rho) &=& \frac{g(\rho)}{\sqrt{f(\rho) +g(\rho) \varphi^2_\rho}}= \frac{g(\rho)}{[g(\rho)-g_0]}~ b_\varphi  \cr
 & & \cr
v_\varphi (\rho) &=& 0 \cr
v_\theta (\rho) &=& \frac{g(\rho)\varphi^2_\rho}{\sqrt{f(\rho) +g(\rho) \varphi^2_\rho}} =\sqrt{\frac{f(\rho)}{g(\rho)}}\frac{g_0}{\sqrt{g(\rho)-g_0}}.
\label{a,b-equations}
\eea
Separating in the $x^\mu$ and $\rho$ variables gives the equations $\partial^2 \Delta\varphi = M^2_\varphi \Delta\varphi$ and $\partial^2 \Delta\theta = M^2_\varphi \Delta\theta$ where $\partial^2$ is the four-dimensional D'Alembertian, while the fluctuation equations in $\rho$ take the form
\be
\frac{\partial}{\partial \rho}\left( b(\rho) \frac{\partial \psi (\rho)}{\partial \rho}  \right) + a(\rho)M^2 \psi (\rho) -v (\rho) \psi (\rho) =0  ,
\label{psi-eq}
\ee
where $\psi (\rho)$ is either  the $\Delta\varphi$ or $\Delta\theta$ with their corresponding $a$, $b$, $M^2$ and $v$ functions.

The $D7$ and $\overline{D7}$ branes stack separately on each side of the $U$-shaped embedding profile.  In order to distinguish these separate locations a transformation of coordinates from $\rho$ to $z$ is introduced with the location of the $D7$ branes along $z>0$ branch while the $\overline{D7}$ branes are along the $z<0$ branch.  The stacks smoothly overlap at $z=0$ corresponding to the location $\rho =\rho_0$ with $\rho_0$ an additional parameter of the model.  A useful transformation of coordinates that separately describes the branches and at the same time converts the fluctuation equations to the form of one-dimensional Schr\"odinger equations \cite{Anguelova:2011bc} is given by
\be
z=\pm\int_{\rho_0}^\rho  d\rho^\prime  \sqrt{\frac{a_\theta(\rho^\prime)}{b_\theta(\rho^\prime)}} =\pm\int_{\rho_0}^\rho  d\rho^\prime  \sqrt{\frac{a_\varphi(\rho^\prime)}{b_\varphi(\rho^\prime)}}  .
\label{zcoord}
\ee
The $D7$ branes lie along positive $z$ while the $\overline{D7}$ branes lie along negative $z$ and the overlap of the 
$D7$ and $\overline{D7}$ branes occurs at the turning point of the $U$-shaped embedding, that is at $\rho = \rho_0$ ($z=0$).  $\rho_0$ is an arbitrary parameter of the model along with $C$ and $T$.  Scaling each fluctuation function by its respective $a$ and $b$ so that
\be 
\psi = \frac{\phi}{(ab)^{1/4}} 
\ee
and substituting into the Euler-Lagrange equation (\ref{psi-eq}) yields the fluctuation equation of the Schr\"odinger  form
\be
-\phi^{\prime\prime} + {\cal V} \phi = M^2 \phi ,
\label{Schrodinger1}
\ee
with 
\be
{\cal V} = v/a + W^{\prime} +W^2 
\ee
where the prime indicates differentiation with respect to $z$ while the superpotential $W$ is given in terms of the prepotential scaling function $U$
\be
U= \ln{(ab)^{1/4}} = \frac{1}{4} \ln{ab}  ,
\ee
with
\be
W = U^\prime =\frac{1}{4} \frac{(ab)^\prime}{ab} .
\ee
Inverting the coordinate transformation to find $\rho =\rho (z)$, the potential ${\cal V}$ is now expressed in terms of the $z$ coordinate, ${\cal V}(\rho (z))$.  

The fluctuation spectrum values of $M^2$ for the techni-scalar and techni-pseudoscalar meson modes must in general be determined numerically for various values of $\rho_0$, $C$ and $T$.  Insight into the spectrum can be obtained by considering values of $\rho$ in the walking region where the metric simplifies and $f$ and $g$ are approximated by equation (\ref{walkingfandg}).  From equations (\ref{a,b-equations}) the simplified ratio $a/b$ is found to be
\be
a/b= 4CT \frac{e^{4\rho}}{(1-e^{4(\rho_0-\rho)})},
\ee
resulting in the coordinate transformation
\be
z =\pm \sqrt{CT} \sqrt{e^{4\rho} -e^{4\rho_0}} .
\ee
Scaling out the $\sqrt{CT e^{4\rho_0}}$ factors and introducing the new variable $\zeta =\frac{e^{-2\rho_0}}{\sqrt{CT}}~z  = \pm \sqrt{e^{4(\rho - \rho_0)}-1}$, the prepotentials for the $\Delta\varphi$ and $\Delta\theta$ fluctuations are obtained from the respective products of $a$ and $b$ as
\bea
U_{\varphi}&=& \frac{1}{4} \ln{a_\varphi b_\varphi}=\frac{1}{4}\ln{\left[\frac{1}{16^3 C T e^{-4\rho_0}}\left(\frac{\zeta^4}{\zeta^2 + 1} \right) \right] }\cr
U_{\theta}&=& \frac{1}{4} \ln{a_\theta b_\theta}=\frac{1}{4}\ln{\left[\frac{1}{16^3 C T e^{-4\rho_0}}\left(\zeta^2 + 1\right)\right] }.
\eea

The superpotentials follow directly by differentiating with respect to $z$ where now the prime stands for differentiation with respect to $\zeta$ so that
\bea
W_\varphi &=&\frac{d\zeta}{dz}U_\varphi^\prime = \frac{1}{\sqrt{CTe^{4\rho_0}}} \left[ \frac{1}{\zeta} -\frac{1}{2} \frac{\zeta}{(\zeta^2 +1)} \right]  \cr
W_\theta &=&\frac{d\zeta}{dz}U_\theta^\prime = \frac{1}{\sqrt{CTe^{4\rho_0}}} \left[ \frac{1}{2} \frac{\zeta}{(\zeta^2 +1)} \right] .
\eea
The superpotential for the $\Delta\varphi$ fluctuations is singular \cite{Panigrahi:1993zy}, \cite{Cooper:2001zd}.  Hence the positivity of the energy (mass squared) cannot be proven from the SUSY (Riccati) form of the Hamiltonian.  The singularity in the superpotential maps the energy eigenfunctions, $\phi_n$, out of the Hilbert space, so that  $|| (W\pm ip)\phi_n ||$ diverges even though $||\phi_n ||$ is finite.  The non-vanishing explicit potential $v_\theta$ breaks the SUSY albeit with a very small coefficient for the chosen values of $T$ and $\rho_0$.  Ignoring the SUSY breaking the $\Delta\theta$ superpotential is non-singular, however it does not vanish quickly enough at large $\zeta$ for there to be a normalizable ground state with zero energy.  (For the above prepotentials and their SUSY partner potentials, the zero energy ground states are of the form $\phi_0 =e^{\pm U}= ab^{\pm 1/4}$ which, for large $\zeta$, go as  $\zeta^{\pm 1/2}$ and are not normalizable).  Since the walking region is of finite length the potentials will be cut-off at large $\rho=\rho_\Lambda <\rho_*$.  That is there are infinite walls at $\rho = \rho_\Lambda$.  Hence the energy eigenfunctions must vanish at the walls and the above \lq\lq would be zero energy" ground state wavefunction does not satisfy this boundary condition.  

\begin{figure*}
\begin{center}
$\begin{array}{cc}
\includegraphics[scale=0.70]{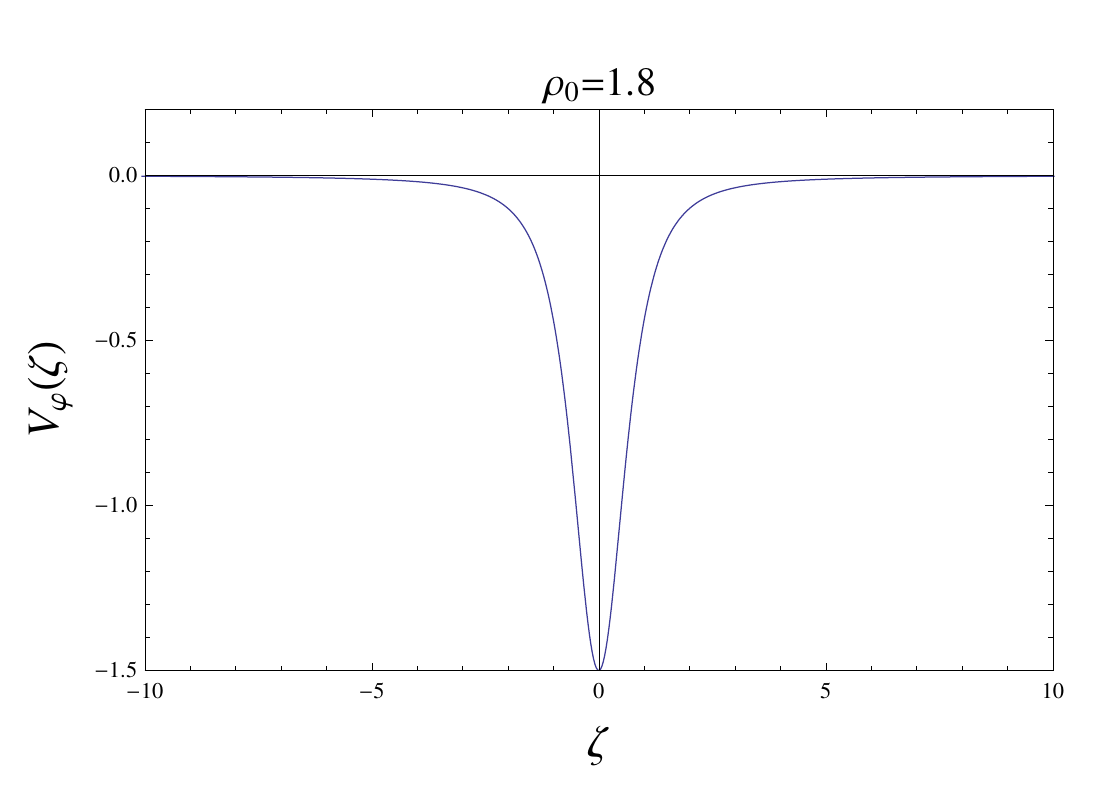} &
\includegraphics[scale=0.70]{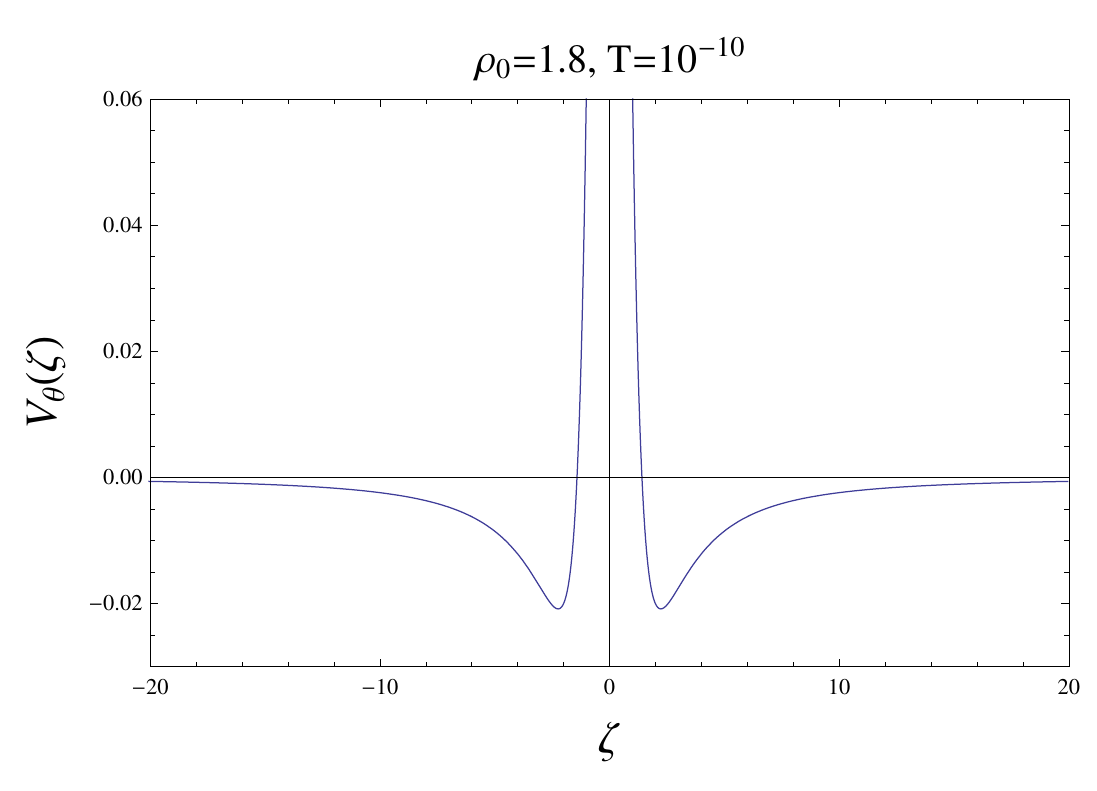}\\
\end{array}$
\caption{The scaled coordinate walking region potentials for $\Delta\varphi$, $V_\varphi$ on the left with minimum value -3/2, and for $\Delta\theta$, $V_\theta$ on the right with maximum value 1/2.}
\label{fig:WalkingPotentials}
\end{center}
\end{figure*}

In addition to the superpotential, the potential also includes supersymmetry breaking pieces for the $\Delta\theta$ fluctuations.  That is, while the explicit SUSY breaking potential term for $\Delta\varphi$ vanishes, $v_\varphi =0$, it follows from equation (\ref{a,b-equations}) that
\be
v_\theta /a_\theta = \left( T e^{4\rho_0} \right) \frac{1}{z^2 +CT e^{4\rho_0}}= \left( \frac{T e^{4\rho_0}}{CT e^{4\rho_0}} \right) \frac{1}{(\zeta^2 + 1)} .
\ee
Differentiating the superpotentials, the potentials for each fluctuation are found to be
\bea
{\cal V}_{\varphi}&=&\frac{d\zeta}{dz} W_\varphi^{\prime} +W_\varphi^2 =\left(\frac{1}{CTe^{4\rho_0}}\right)\left[W_\varphi^\prime + W_\varphi^2 \right] \cr
 &=&-\left( \frac{1}{CTe^{4\rho_0}}\right) \frac{1}{4}\frac{\left( \zeta^2 +6\right)}{\left( \zeta^2 + 1 \right)^2}  \cr
 &\equiv&\left( \frac{1}{CTe^{4\rho_0}}\right) V_\varphi (\zeta) \cr
 & & \cr
{\cal V}_{\theta}&=& v_\theta /a_\theta + \frac{d\zeta}{dz} W_\theta^{\prime} +W_\theta^2 = \left(\frac{1}{CTe^{4\rho_0}}\right) \left[ \frac{1}{(\zeta^2 +1)}+ W_\theta^\prime +W_\theta^2 \right] \cr
 &=& -\left( \frac{1}{CTe^{4\rho_0}}\right) \left\{\frac{1}{4}\frac{\left( \zeta^2 -2 \right)}{\left( \zeta^2 + 1 \right)^2}+ T e^{4\rho_0}\frac{1}{\left( \zeta^2 + 1 \right)}\right\} \cr
 &\equiv&\left( \frac{1}{CTe^{4\rho_0}}\right) V_\theta (\zeta) .
\eea
Finally multiplying equation (\ref{Schrodinger1}) by $(CTe^{4\rho_0})$ and expressing all in terms of the variable $\zeta$, the Schr\"odinger equations for the fluctuations take the form
\bea
-\phi_\varphi^{\prime\prime} (\zeta) + V_\varphi (\zeta) \phi_\varphi (\zeta) = m^2_\varphi \phi_\varphi (\zeta) \cr
-\phi_\theta^{\prime\prime} (\zeta) + V_\theta (\zeta) \phi_\theta (\zeta) = m^2_\theta \phi_\theta (\zeta) ,
\eea
where the prime here again indicates differentiation with respect to $\zeta$ and the squared masses have been scaled so that
\be
m_\varphi^2 = CT M_\varphi^2 e^{4\rho_0} \qquad, \qquad m_\theta^2 = CT M_\theta^2 e^{4\rho_0} .
\label{scalemasses}
\ee
Finally, the scaled potentials take the form
\bea
V_\varphi (\zeta) &=& -\frac{1}{4}\frac{\left( \zeta^2 +6 \right)}{\left( \zeta^2 + 1\right)^2}  \cr
V_\theta (\zeta) &=& -\frac{1}{4}\frac{\left( \zeta^2 -2 \right)}{\left( \zeta^2 + 1 \right)^2}-T e^{4\rho_0}\frac{1}{\left( \zeta^2 + 1 \right)} 
\label{scaledpotentials}
\eea
and are plotted in Fig.(\ref{fig:WalkingPotentials}).  This agrees with the potentials given in reference \cite{Anguelova:2012ka}. For the small values of $T$ necessary for a walking region of reasonable $\rho$ length, the second term in $V_\theta$ is small compared to the first term. However it nonetheless plays a pivotal role in the determination of the spectrum as shown in detail in the Appendix.  Note that the value of $V_\theta$ at $\zeta =0$ is $V_\theta (0) =1/2 -Te^{4\rho_0} \approx 1/2$ while $V_\varphi (0) =-3/2$.

Although the $\Delta\varphi$ superpotential is singular, its potential is non-singular.  The spectrum for this Schr\"odinger equation has a negative energy ground state since $\int d\zeta V_\varphi (\zeta) \leq 0$ and  $\int d\zeta |\zeta|^\gamma |V_\varphi (\zeta)| < \infty$, for $0 \leq \gamma <1$ \cite{Blankenbecler:1977pf,Kocher}.  Indeed the $\Delta\varphi$ potential obeys these conditions with
\be
\int_{-\infty}^\infty d\zeta V_\varphi (\zeta) = -\frac{7\pi}{8}.
\ee
Introducing a UV cut-off so that as the walking region boundary walls move in towards the main region of support of the $\Delta\varphi$ potential the contribution of the infinite walls to the ground state energy  becomes more significant and the mass squared will be positive indicating stability of the brane's oscillation.  As detailed in the Appendix, this stability crossing occurs at approximately the equality of the square well energy, $\pi^2/(4 \zeta_\Lambda^2)$, to that of the absolute value of the ground state energy without walls, which is numerically determined as 0.56.  This yields a cut-off value $\zeta_\Lambda = 2.10$.  Thus the scalar $\Delta\varphi$ fluctuations are unstable for walking regions of length greater than $\rho_\Lambda - \rho_0 = 0.30$.

The numerical determination of the brane oscillation stability is obtained as follows. The distance of walking from the $D7$-$\overline{D7}$ overlap at $\rho =\rho_0$ to $\rho <\rho_*$  is related to the final renormalization scale $\zeta$ according to $\rho -\rho_0 = \frac{1}{4} \ln{[\zeta^2 +1]}$.  Using the naive ending of the walking region as $\rho_*$ the corresponding scaled coodinate is $\zeta_* = \sqrt{[e^{4(\rho_* -\rho_0)}-1]}$.  For $\rho_0 =1.8$, the $T=10^{-10}$ case yields $\rho_* = 5.66$ corresponding to $\zeta_* = 2,252$ while the $T=10^{-12}$ case yields $\rho_* = 6.81$ corresponding to $\zeta_* =22,470$.  Both values are far beyond the location of the extrema of the potentials.  Thus it is reasonable to cutoff the potentials after a modest walking length as the wavefunction $\phi$ of any negative energy bound state ($m^2<0$) will be highly localized about the origin as are the potentials.  In addition the extended technicolor physics is expected to be present at high energy in the walking region, so the model, to include this new physics, must be altered.  Hence a high momentum cut-off will be used to indicate the necessity of this new physics entering the model \cite{Anguelova:2011bc}.  

\begin{figure*}
\begin{center}
$\begin{array}{cc}
\includegraphics[scale=0.79]{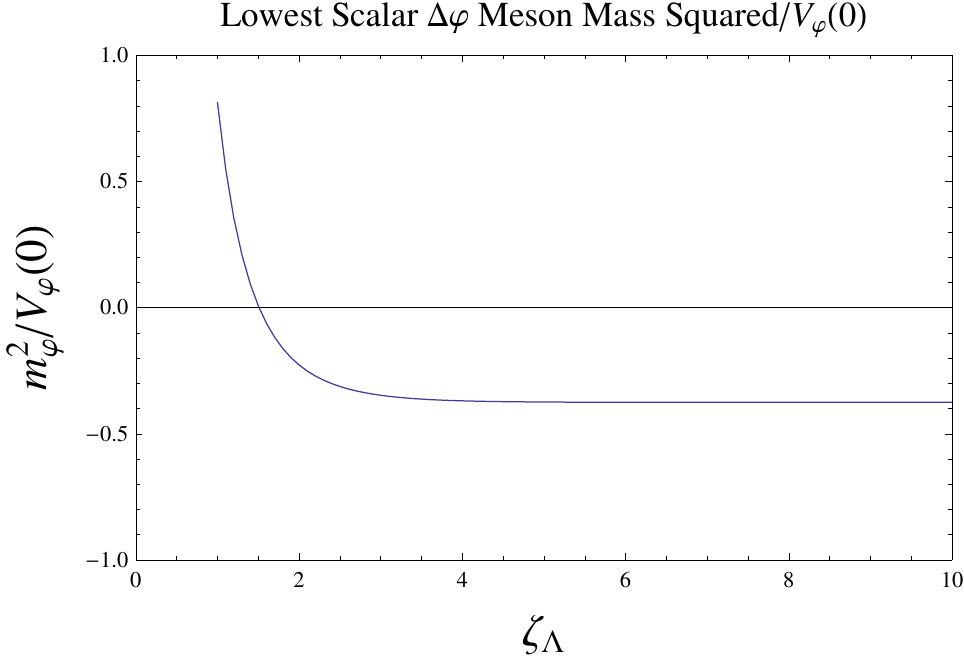} &
\includegraphics[scale=0.70]{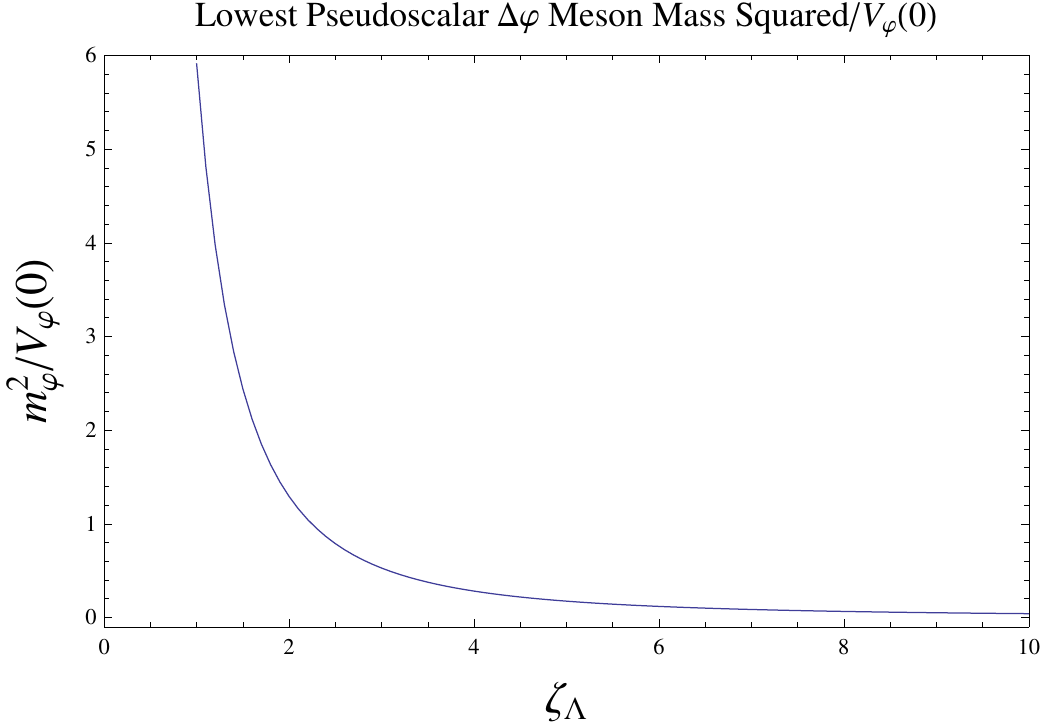}\\
\includegraphics[scale=0.62]{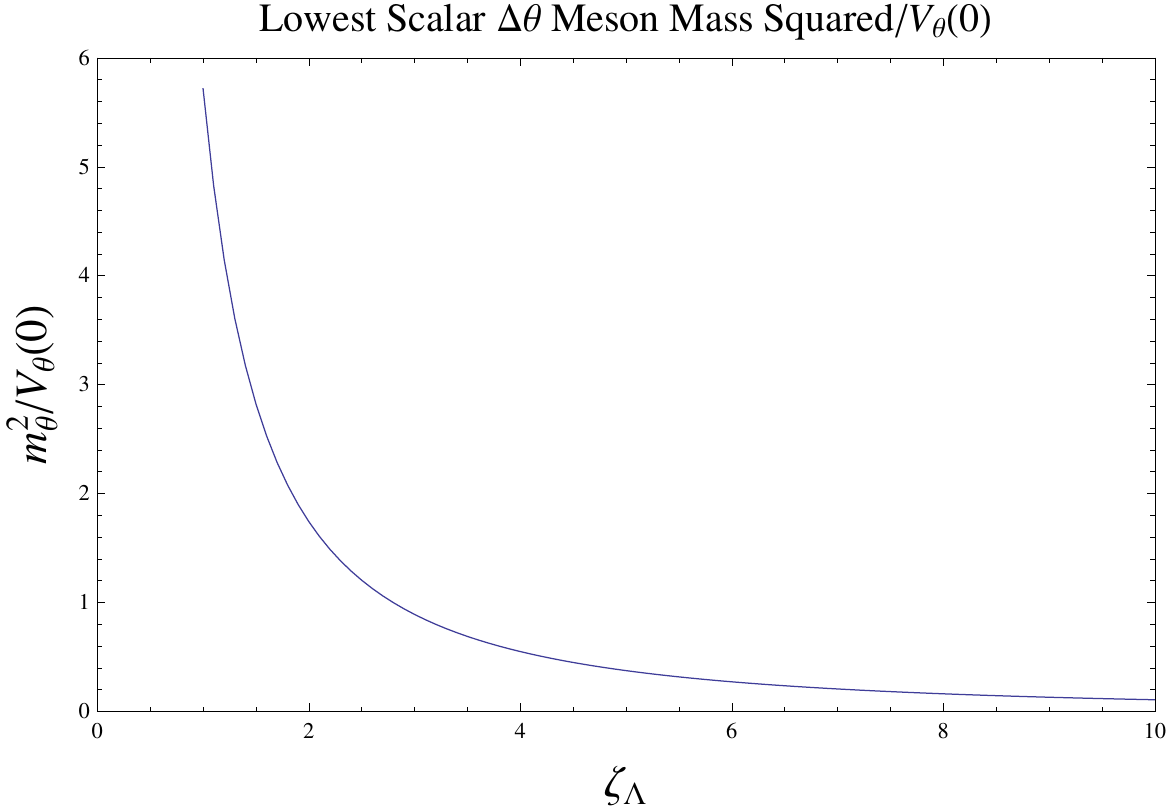} &
\includegraphics[scale=0.65]{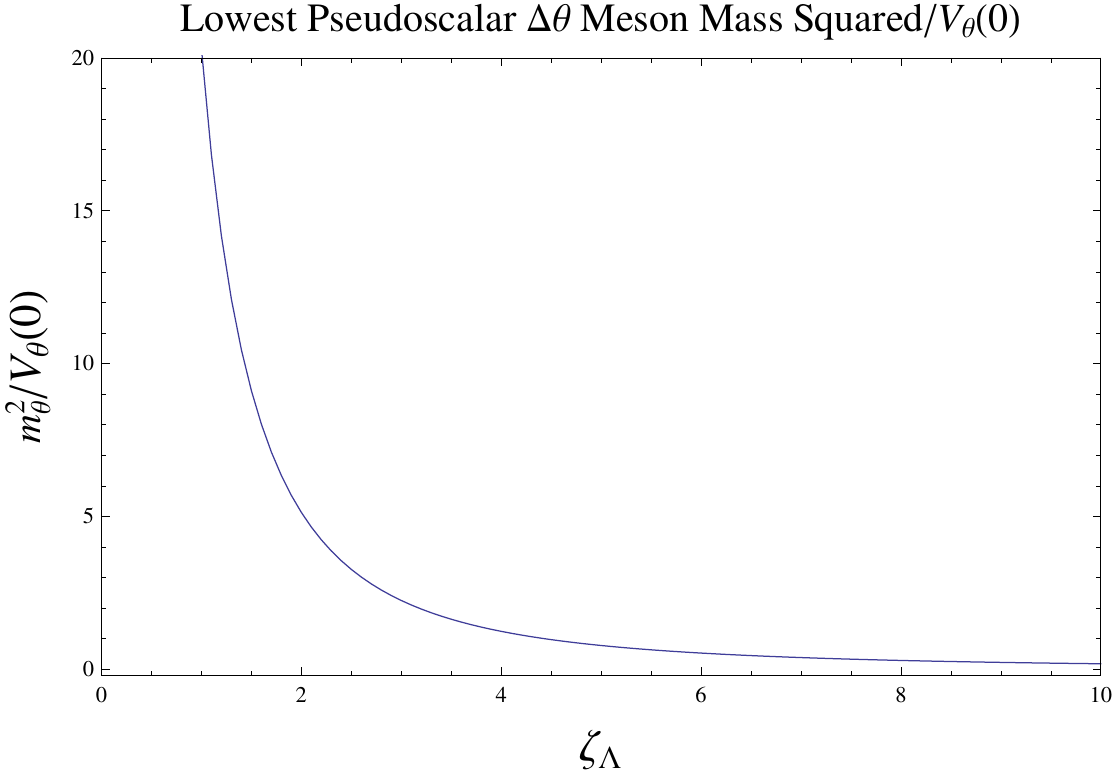}
\end{array}$
\caption{Lowest values of the $\Delta\varphi$ squared masses divided by the potential at the origin, $m_\varphi^2 / |V_\varphi (0)|$, and $\Delta\theta$ squared masses divided by the potential at the origin, $m_\theta^2 /|V_\theta (0)|$, versus the scaled coordinate cut-off value $\zeta_\Lambda$ for the scalar and pseudoscalar mesons.  The $D7$-$\overline{D7}$ overlap position $\rho_0 =1.8$ for all plots in the panel while the $\Delta\theta$ potential includes the $v_\theta/a_\theta$ potential with $T=10^{-10}$}.
\label{fig:WalkingMassSquared}
\end{center}
\end{figure*}

Alternatively stated, the spectrum of $D7$-$\overline{D7}$ lowest oscillation frequencies will be obtained as the cut-off is varied.  Denoting the cut-off as $\zeta_\Lambda$ or correspondingly $\rho_\Lambda =\rho_0 +\frac{1}{4}\ln{[\zeta^2_\Lambda +1]}$, the Schr\"odinger equations for the fluctuations are solved for scalar and pseudoscalar meson boundary conditions at the origin.  Specifically, scalar meson fluctuations are even functions of $\zeta$ such that the derivative of their wavefunction vanishes at the origin, $\phi^\prime (\zeta =0) =0$.  For these linearized scalar oscillations the wavefunction is normalized to 1 at the origin as well, $\phi (\zeta=0)=1$.  The value of the lowest oscillation frequency, $m_\varphi^2$ or $m_\theta^2$, is then tuned so that the wavefunction first vanishes at the cut-off, $\phi (\zeta_\Lambda )=0$.  If this is negative, the oscillation grows exponentially in time, hence the fluctuation is not stable.  Likewise, the pseudoscalar meson fluctuations are odd functions of $\zeta$ so that their wavefunction vanishes at the origin, $\phi (\zeta=0)=0$.  Furthermore, the derivative of the wavefunction is normalized to 1 at the origin, $\phi^\prime (\zeta =0) =1$.  As in the scalar meson case, the value of the lowest oscillation frequency, $m_\varphi^2$ or $m_\theta^2$, is then tuned so that the wavefunction first vanishes (beyond the origin) at the cut-off, $\phi (\zeta_\Lambda )=0$.  If this is negative, the oscillation is growing exponentially in time, hence the fluctuation is not stable.  These lowest values of the mass squared for the scalar and pseudoscalar $\Delta\varphi$ and $\Delta\theta$ mesons divided by the respective potentials at the origin, $|V_\varphi (0)|=3/2$ and $|V_\theta (0)|=1/2$, are plotted as a function of the cut-off $\zeta_\Lambda$ in Fig.( \ref{fig:WalkingMassSquared}).  As can be seen, the $\Delta\varphi$ scalar meson is stable only for cut-off values below $\zeta_\Lambda = 1.5$ (corresponding to $\rho$ values above $\rho_0 =1.8$ but below the cut-off value of $\rho_\Lambda = 2.095$).  In such a case, the cut-off confines the wavefunction to be close to the origin.  As the cut-off potential well walls move out beyond the majority of support of the potential $V_\varphi$ there effect becomes irrelevant as the wavefunction is highly localized around the origin and the potential exhibits one negative energy bound state with asymptotic energy $m_\varphi^2 \approx -0.563$ ($m_\varphi^2 /|V_\varphi (0)|\approx -0.375$).  On the other hand the $\Delta\varphi$ pseudoscalar meson and both $\Delta\theta$ scalar and pseudoscalar meson fluctuations are stable.  As the cut-off grows, the mass squared values are tuned to smaller and smaller positive values for a particle in an infinite well.  As the end of the walking region is approached the validity of the simplified walking region metric comes into question and the cut-off must be kept less than the walking region limit $\rho_\Lambda < \rho_*$.  

To summarize the case for the scalar $\Delta\varphi$ meson, when the cut-off is of order 1, the spectrum is dominated by the contribution arising from the infinitely high walls so  that the oscillation frequencies are real and the brane stable.  Such a situation corresponds to the gauge coupling exhibiting a walking behavior only over a comparatively very small range. As the well walls move out past the main support of the potential, the infinite well eigenenergy becomes smaller and smaller so that the squared mass is dominated by the potential's bound state energy.  This is negative and approaches a constant value as the cut-off is removed (at least as far as the walking region cut-off $\rho_\Lambda$).  The $D7$ brane exhibits an unstable oscillation mode growing exponentially in time.  

As detailed in the Appendix, the potential cannot overcome the infinite well energy for the pseudoscalar $\Delta\varphi$ vibration even as the cut-off is removed the squared masses are positive, approaching zero from above.  The potential for the $\Delta\theta$ scalar and pseudoscalar meson modes leads to an infinite tower of bound states as given by equation (42) in the absence of a cut-off.  The eigenvalues are small due to the size of $T e^{4\rho_0}<<1$.  As the infinite well walls move in towards the origin, its positive contribution to the energy eigenvalue dominates that of the $\Delta\theta$ well.  This brane oscillation mode becomes stable at approximately where the energies cross, given by equation (44) with $\lambda = T e^{4\rho_0}\approx 1.34 \times 10^{-7}$ which leads to an extremely large value for the cut-off, $\rho_\Lambda \geq 2,000$.  For small cut-off values, $\zeta_\Lambda \sim O(10)$ corresponding to $\rho_\Lambda \sim 3$, the $D7$ brane $\Delta\theta$  oscillation modes are stable.

\begin{figure*}
\begin{center}
$\begin{array}{cc}
\includegraphics[scale=1.0]{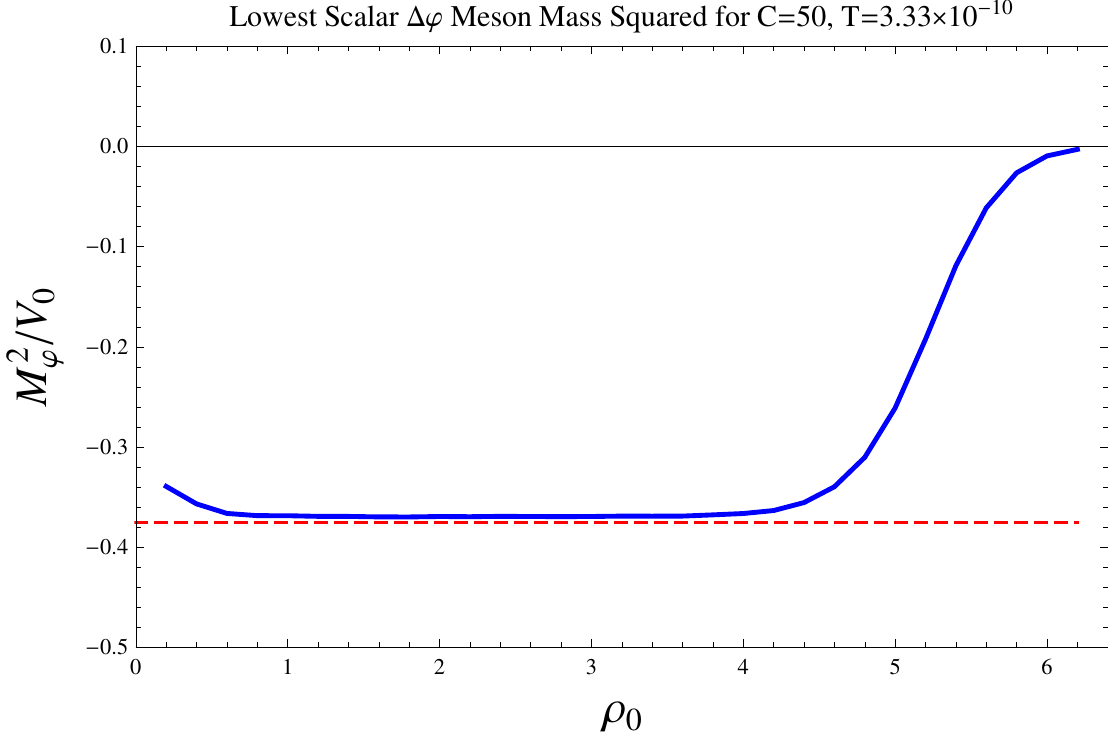}
\end{array}$
\caption{The solid blue curve depicts the values of the smallest techni-brane oscillation frequencies squared divided by the potential at the origin, $M_\varphi^2 /|{\cal V}_\varphi (z=0)|$, as a function of the $D7$-$\overline{D7}$ overlap position $\rho_0$ for fixed parameters $C=50$ and $T=3.33\times 10^{-10}$.  The dashed red curve indicates the squared frequencies divided by the potential at the origin, $3 CT M_\varphi^2 e^{4\rho_0}/2$, obtained for $\rho_0$ in the walking region for large cut-off (as given in the upper left panel of Fig.( \ref{fig:WalkingMassSquared})).  The negative values indicate that the lowest frequency $\Delta\varphi$ oscillation mode is unstable.}
\label{fig:ExactScalarMasses}
\end{center}
\end{figure*}

To remove the cut-off completely, or for cut-off values beyond $\rho_*$, the full metric and corresponding functions $f$ and $g$ of equation (\ref{fullfandg}) must be used.  This requires a full numerical analysis.  Returning to the expressions for $f$ and $g$ in equation (\ref{fullfandg}), the $\Delta\varphi$ fluctuation equation is transformed to a Schr\"odinger equation form using equation (\ref{a,b-equations}) for $a_\varphi$ and $b_\varphi$.  The change of coordinates to $z$ in equation (\ref{zcoord}) must be numerically evaluated as well as inverted numerically to find $\rho =\rho (z)$.  The case of fixed parameters $C=50$ and $T=3.33\times 10^{-10}$ is studied for the $D7$-$\overline{D7}$ overlap location $0.2\leq \rho_0 \leq 6.2$ with no cut-off.  The potential ${\cal V}_\varphi =W^2_\varphi + W_\varphi^\prime$ is digitized as a function of $z$. As in the walking region case, the mass squared can be scaled by the value of the potential ${\cal V}_\varphi$ at the origin for each choice of $\rho_0$, the value $|{\cal V}_\varphi (z=0)|\equiv V_0$.  In the walking region this was the scaled mass squared factor $3 CT M_\varphi^2 e^{4\rho_0}/2$ as plotted in the upper left panel of Fig.( \ref{fig:WalkingMassSquared}).  The vibration modes were then obtained by numerically integrating the Schr\"odinger equation while tuning the squared mass parameter to yield a normalizable wavefunction for the case of no cut-off.  The smallest squared mass eigenvalues, scaled by the minimum of the ${\cal V}_\varphi$ potential, $M^2_\varphi/V_0$, for the $\Delta\varphi$ scalar meson case are shown in Fig.( \ref{fig:ExactScalarMasses}).  The negative eigenvalue is in agreement with the approximate results obtained from the walking region as detailed in the Appendix.  Regardless of the $D7$-$\overline{D7}$ overlap location, the scalar $\Delta\varphi$ meson oscillation is unstable.

\section{Conclusion}

The gravity dual to a ${\cal N}=1$ SUSY gauge theory which exhibits an approximately conformal (\lq\lq walking") region while strongly interacting was previously introduced \cite{Nunez:2008wi}.  Stacks of $D7$-$\overline{D7}$ techni-branes corresponding to the addition of techni-fermions in the gauge theory were embedded in the 10 dimensional space-time as prescribed in \cite{Anguelova:2011bc}.  Fluctuations of the embedded branes into complementary ($\varphi$ and $\theta$) space were considered \cite{Anguelova:2012ka}.  The linearized Euler-Lagrange equations (\ref{psi-eq}) were obtained for oscillations into the complementary space as functions of the bulk coordinate $\rho$.  A coordinate transformation (\ref{zcoord}) then resulted in the equations of motion \cite{Anguelova:2011bc,Anguelova:2012ka} taking the form of  one-dimensional Schr\"odinger equations (\ref{Schrodinger}) with superpotentials $W=dU/dz$ obtained from the prepotential scaling function $U=\ln{(ab)^{1/4}}$.  

The scalar and pseudoscalar meson mass squared spectrum was obtained by numerically analyzing the Schr\"odinger equations.  With a bulk coordinate cut-off taken to be in the walking region only, $1\leq\rho \leq \rho_\Lambda< \rho_*$, the fluctuation potentials were found explicitly, (cf. Eq. (\ref{scaledpotentials})), and plotted in Fig.(\ref{fig:WalkingPotentials}).  The  $\Delta\theta$ meson modes form an infinite tower of negative states in the absence of cut-off walls. As the walls are brought in from infinity these energies however become positive when the contribution of the square well wall's energy exceeds the potential's bound state energy. This occurs at extremely large values of the cut-off due to the smallness of $T$. Correspondingly the brane oscillations are stable  as depicted in the two lower panels in Fig.( \ref{fig:WalkingMassSquared}).  As seen in the Appendix, the mass squared eigenvalue for the $\Delta\varphi$ pseudoscalar meson vanishes in the SUSY ($\lambda \rightarrow 0$) limit with no cut-off.  When the cut-off is introduced all energies are  positive, as depicted in the upper right panel in Fig.(\ref{fig:WalkingMassSquared}), and hence exhibit stable oscillations.  

On the other hand, the $\Delta\varphi$ scalar meson ground state energy arises from two sources. One is the negative energy contribution arising from the localized potential and the other is a positive term coming from the infinitely high walls from the cut-off. Unless the walls are very close to the position of the highly localized potential, the negative energy contribution of the potential dominates. Only for small values ($< \zeta_\Lambda=2.095$) of the cut-off does the positive energy contribution of the walls takes over leading to a stable brane. As the cut-off increases, the positive energy contribution of the infinite walls decreases and the negative energy of the brane dominates the fluctuation mass squared eigenvalue, hence destabilizing the oscillations of the brane.  As the cut-off walls move beyond the major support of the brane's potential, the square well contribution to the mass squared energy is negligible.
The brane's wavefunction is highly localized in this bound state about the origin, i.e. close to $\rho_0$.  The negative mass squared then attains its constant value as seen in the upper left panel of Fig.( \ref{fig:WalkingMassSquared}).  For the no cut-off case, the full metric must be used not just its simplified form in the walking region to find the Schr\"odinger equation potentials.  The numerical analysis for the $\Delta\varphi$ scalar meson fluctuation mass squared eigenvalue was performed as a function of the $D7$-$\overline{D7}$ branes overlap position $\rho_0$.  The brane oscillations are unstable.  In the walking region the scalar meson mass squared value agreed with that calculated using the simplified walking region metric as depicted in Fig.(\ref{fig:ExactScalarMasses}).  

The range of stability of the techni-brane oscillations with a cut-off in the walking region is limited to $1.8\leq \rho \leq 2.095$.  Whether a different embedding of the stacks of $D7$-$\overline{D7}$ branes involving the 2-cycle $\Sigma_2$ used to wrap the stacked $D5$ branes of the gauge theory made from the twisting of the $S^2$ coordinates $\theta$, $\varphi$ and the $S^3$ coordinates $\tilde{\theta}$, $\tilde{\varphi}$ and $\psi$ will lead to a stable techni-brane embedding remains to be investigated.

In reference \cite{Anguelova:2012ka}, the authors compared the scalar meson spectrum to that of the vector meson spectrum \cite{Anguelova:2011bc} and concluded the mass splitting was small and hence the brane oscillations were stable.  It has been shown in this paper that the scalar meson spectrum (in particular that of $\Delta\varphi$) is significantly modified compared to the vector meson case.  Indeed as shown above there is a deep lying narrow bound state in the $\Delta\varphi$ scalar meson case with energy given by equation (47) that is not of negligible size. It is the presence of this state, whose ramifications were not properly accounted for previously \cite{Anguelova:2012ka}, which is responsible for the destabilization of the brane except for the case of very small cut-offs.

\begin{acknowledgments}
The authors thank Martin Kruczenski, Georgios Michalogiorgakis and Peter Ouyang for helpful and enlightening discussions.  The work of TEC and STL was supported in part by the U.S. Department of Energy under grant DE-FG02-91ER40681 (Theory).  The work of TtV was supported in part by the NSF under grant PHY-1102585.  
\end{acknowledgments}

\appendix
\section*{Appendix}

\begin{figure*}
\begin{center}
$\begin{array}{cc}
\includegraphics[scale=0.67]{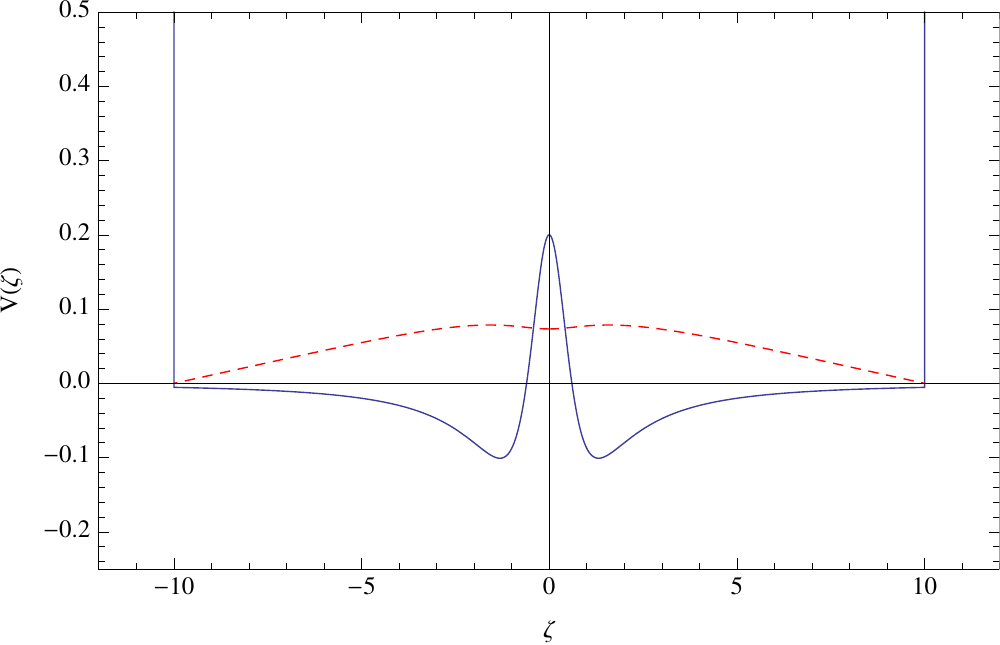} &
\includegraphics[scale=0.7]{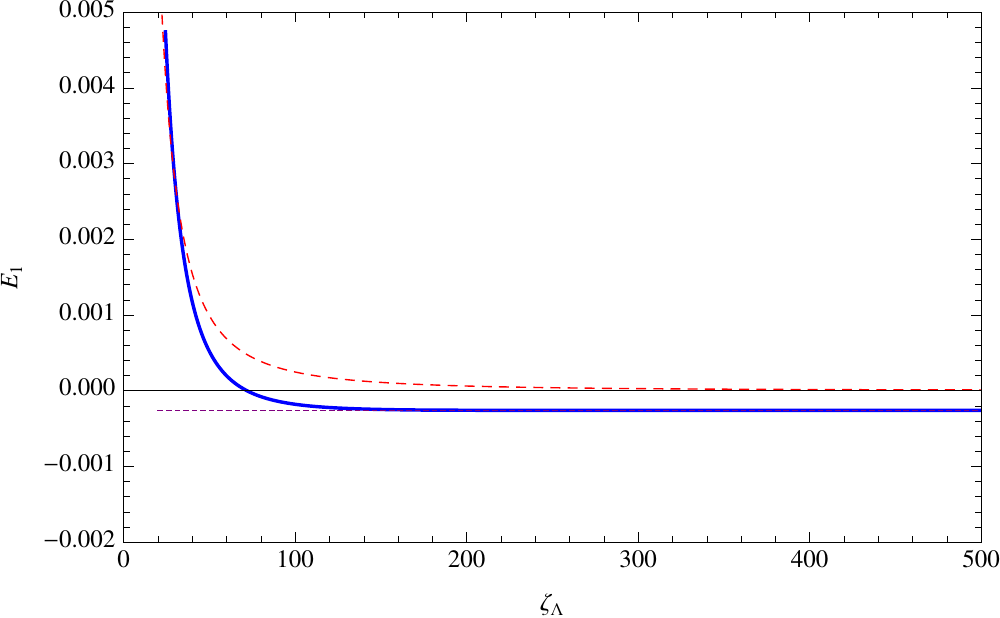}\\
\end{array}$
\caption{Left panel: Model 1 potential as a function of $\zeta$ for cut-off $\zeta_\Lambda=10$ and potential parameter $\lambda=0.1$ (solid blue curve) and corresponding ground state eigenfuction (n=1, dashed red curve). Right panel: ground state energy as a function of the cut-off $\zeta_\Lambda$ for $\lambda=0.1$. Also shown are the asymptotic ground state energy value in the absence of a cut-off (purple dotted line) and the ground state energy of an infinite square well with width equal to two times $\zeta_\Lambda$ and depth equal to zero (red dashed curve).}
\label{fig:cutoff}
\end{center}
\end{figure*}

The structure of the scalar and pseudoscalar mass spectra can be understood in terms of the approximate supersymmetry \cite{Panigrahi:1993zy,Cooper:2001zd}
and conformal symmetry \cite{Case:1950an,Blankenbecler:1977pf}
of equivalent quantum mechanical models with Hamiltonian $H=-\frac{d^2}{d\zeta^2} +V(\zeta)$. \\ \\
{\bf Model 1:}
The normal mode equation describing the  $\Delta \theta$ fluctuations is equivalent to the Schr\"{o}dinger equation of a quantum mechanical model with potential
\begin{eqnarray}
V(\zeta) & = & -\frac{1}{4} \frac{\zeta^2-2}{(\zeta^2+1)^2} - \lambda \frac{1}{\zeta^2+1}, \label{Schrodinger}
\end{eqnarray}
where the potential parameter $\lambda$ is identified with $T e^{4 \rho_0}$ and the energy eigenvalues with the mass-squared values $m_\theta^2$: $E_\theta = m_\theta^2$. The characteristic shape of the potential with a hump at the origin flanked by two valleys is depicted in the left panel of Fig.(\ref{fig:cutoff}). The first term in the Model 1 potential can be written in terms of a superpotential,
\begin{eqnarray}
V(\zeta) & = & W^2(\zeta)+W'(\zeta), \nonumber \\
W(\zeta) & = & \frac{1}{2} \frac{\zeta}{1+\zeta^2}.
\end{eqnarray}
The first term in the Model 1 potential therefore can be viewed as one part of a supersymmetric partner potential system, with the second term providing explicit supersymmetry breaking. Since the superpotential is non-singular, the energy spectrum is necessarily positive-semidefinite \cite{Cooper:2001zd} in the limit of vanishing $\lambda$. 

\begin{figure*}
\begin{center}
$\begin{array}{cc}
\includegraphics[scale=0.68]{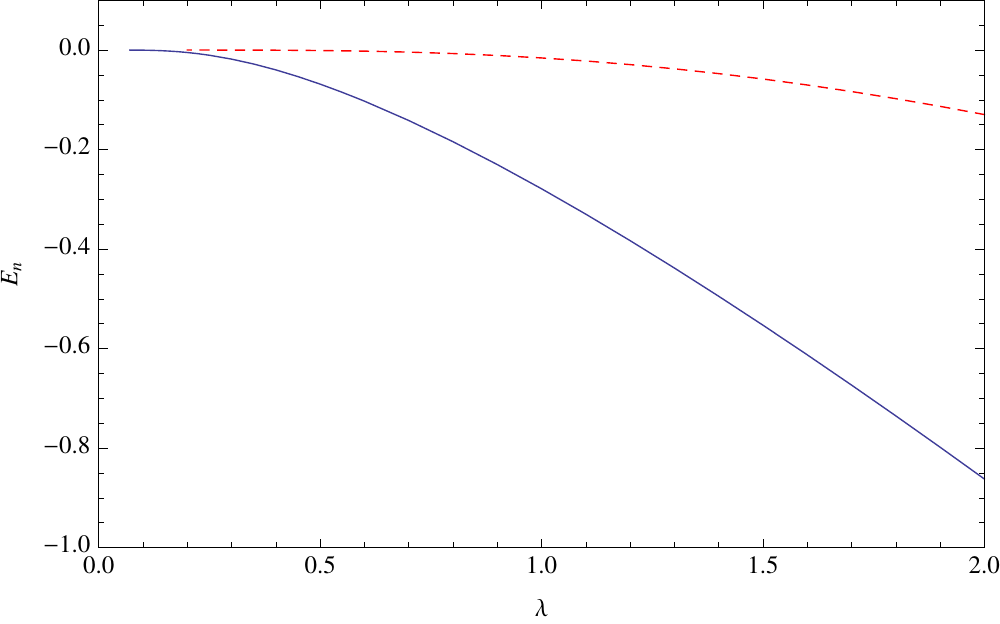} &
\includegraphics[scale=0.70]{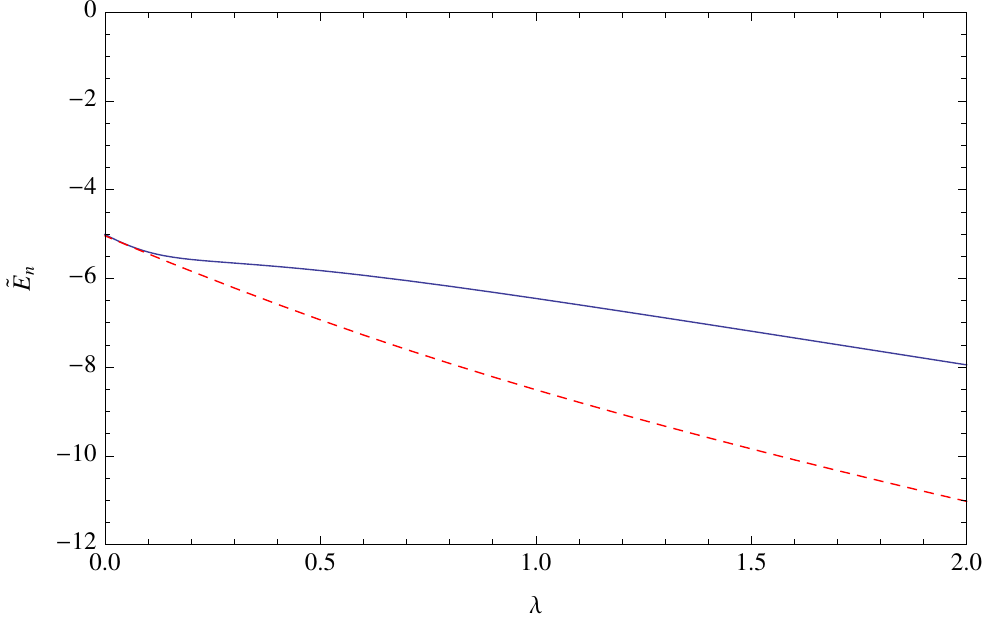}\\
\end{array}$
\caption{Left panel: Model 1 ground state ($n=1$, solid blue curve) and first excited state ($n=2$, dashed red curve) energy $E_n$ as a function of the potential parameter $ \lambda$. Right panel: scaled ground state (n=1, solid blue curve) and first excited state (n=2, dashed red curve) energy $\tilde{E}_n = E_n / e^{- \frac{n \pi}{\sqrt{\lambda}}}$ as a function of $ \lambda$. The convergence of the two curves at small values of $\lambda$ illustrates the emergence of a geometrical pattern in the energy spectrum in this limit.} 
\label{fig:eigenvalues}
\end{center}
\end{figure*}

For  $\zeta \gg 1$ the potential takes the form of the potential of conformal quantum mechanics,
\begin{eqnarray}
V(\zeta) & = & \alpha \frac{1}{\zeta^2}, \label{conformal}
\end{eqnarray}
with $\alpha= -1/4-\lambda$.  Therefore Model 1 can also be viewed as conformal quantum mechanics with  regulated short distance singularity \cite{Case:1950an,Blankenbecler:1977pf}. For $\alpha < -1/4$ such models have a tower of an infinite number of bound states. In the limit of small $\lambda$, it has been numerically verified for low $n$ that the bound state spectrum takes on the characteristic geometrical form
\begin{eqnarray}
E_n & = & - C_1 e^{- \frac{n \pi}{\sqrt{\lambda}}},\,\,\,\,\,\,\,\,\,\,\,\, n \in \mathbb{Z}^+.
\end{eqnarray}
The constant $C_1$ depends on the details of the short distance regularization. For  Model 1 this parameter is numerically determined to take the value $C_1=5.02$. The left panel of Fig.(\ref{fig:eigenvalues}) shows the ground state and first excited state energy levels as a function of $\lambda$. The right panel illustrates the emergence of a geometrical pattern in the energy eigenvalues.

Once a large distance cut-off $\zeta_\Lambda$  is introduced the conformal invariance is further broken. The energy of the negative energy states for positive values of $\lambda$ increases as the cut-off is lowered. The energy of the $n^{\rm th}$ level turns positive approximately when the magnitude of its energy in the absence of a cut-off equals the ground state energy of an infinite square well of width twice $\zeta_\Lambda$ and depth zero, that is, when
\begin{eqnarray}
C_1 e^{- \frac{n \pi}{\sqrt{\lambda}}} & \approx & \frac{\pi^2}{ 4 \zeta_\Lambda^2}.
\end{eqnarray}
In particular,  the ground state energy turns positive for a cut-off value of approximately
\begin{eqnarray}
\zeta_\Lambda & \approx & \frac{\pi}{2 \sqrt{C_1}} e^{\frac{\pi}{2 \sqrt{\lambda}}}\, .
\end{eqnarray}
Below this cut-off value no negative energy states persist in the spectrum. The ground state wave function is displayed in the left panel of Fig.(\ref{fig:cutoff}) for $\zeta_\Lambda= 10$ and $\lambda=0.1$. The right panel shows the dependence of the ground state energy on $\zeta_\Lambda$ for $\lambda=0.1$.\\

\begin{figure*}
\begin{center}
$\begin{array}{cc}
\includegraphics[scale=0.68]{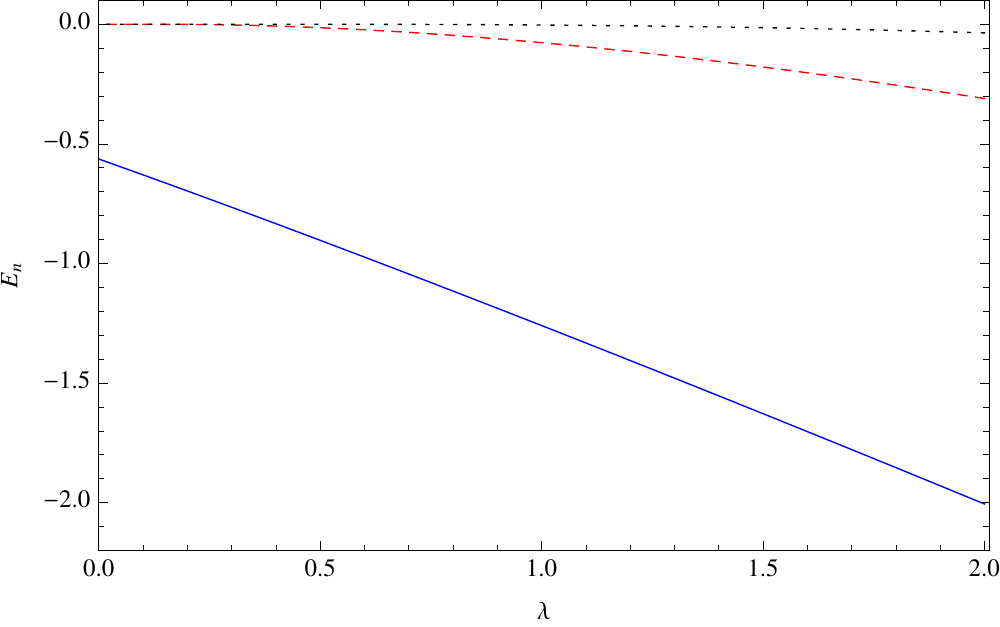} &
\includegraphics[scale=0.70]{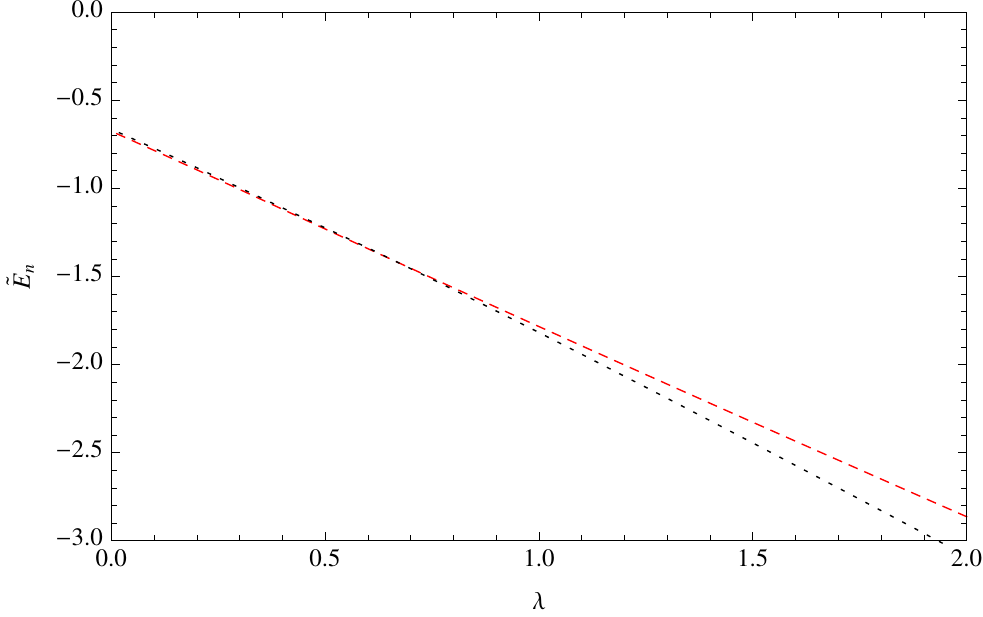}\\
\end{array}$
\caption{Left panel: Model 2 ground state ($n=0$, solid blue curve), first excited state ($n=1$, dashed red curve), and second excited state  ($n=2$, dotted black curve) energy $E_n$ as a function of the potential parameter $ \lambda$. Right panel: scaled first excited state (n=1, dashed red curve) and second excited state (n=2, dashed red curve) energy $\tilde{E}_n = E_n / e^{- \frac{n \pi}{\sqrt{\lambda}}}$ as a function of $ \lambda$. The convergence of the two curves at small values of $\lambda$ illustrates the emergence of a geometrical pattern in one sector of the energy spectrum in this limit.} 
\label{fig:eigenvaluesII}
\end{center}
\end{figure*}

\begin{figure*}
\begin{center}
$\begin{array}{cc}
\includegraphics[scale=0.67]{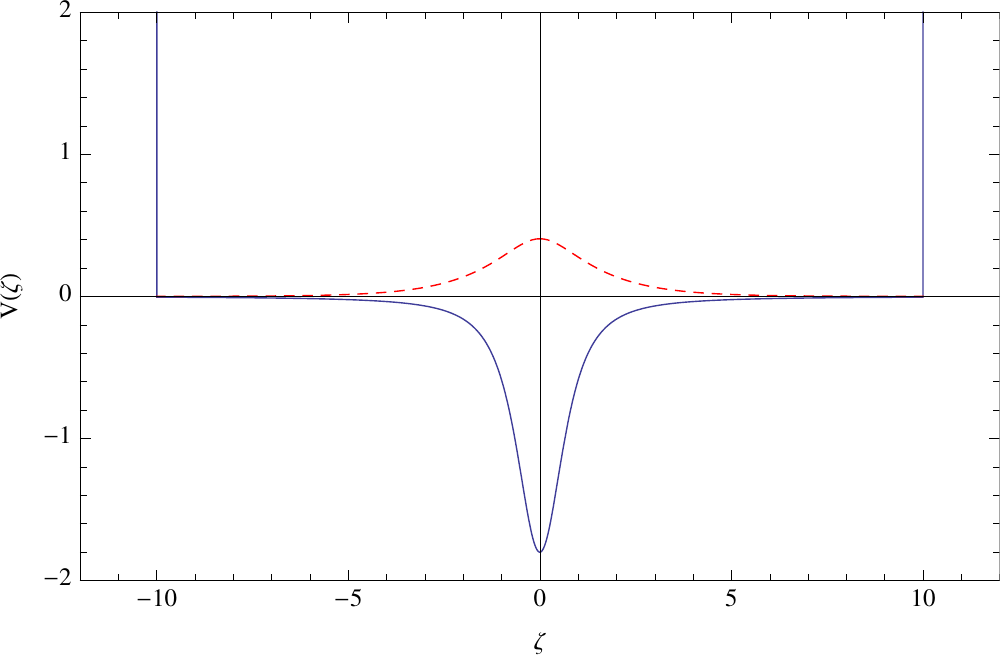} &
\includegraphics[scale=0.7]{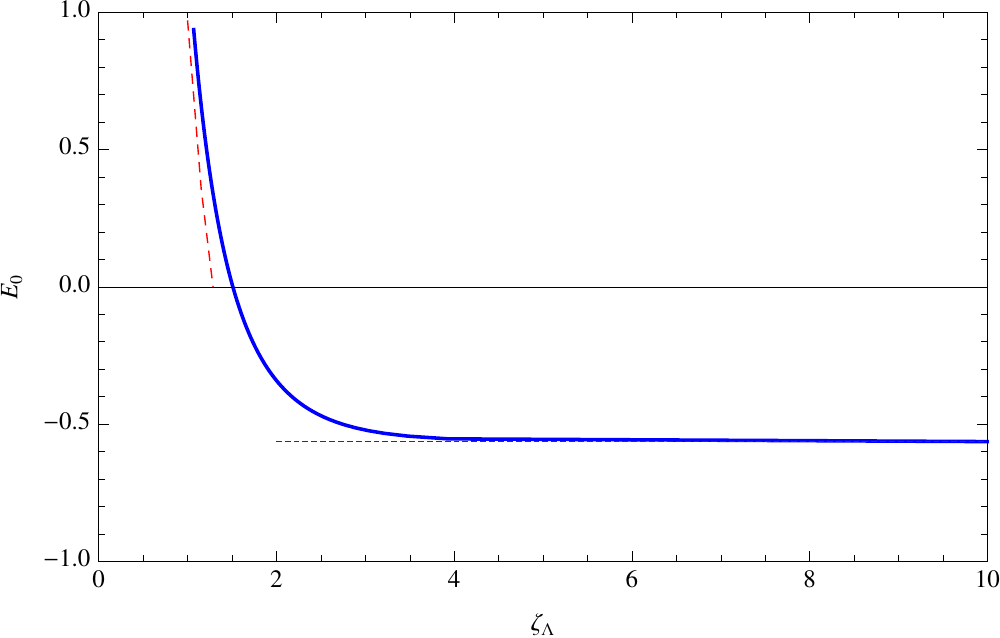}\\
\end{array}$
\caption{Left panel: Model 2 potential as a function of $\zeta$ for cut-off $\zeta_\Lambda=10$ and potential parameter $\lambda=0.1$ (solid blue curve) and corresponding ground state eigenfuction (n=0, dashed red curve). Right panel: ground state energy as a function of the cut-off $\zeta_\Lambda$ for $\lambda=0$. Also shown are the asymptotic ground state energy value in the absence of a cut-off (purple dotted line) and the ground state energy of an infinite square well with width equal to two times $\zeta_\Lambda$ and depth $-3/2$ (red dashed curve).}
\label{fig:cutoffII}
\end{center}
\end{figure*}

{\bf Model 2:}
The normal mode equation describing the  $\Delta \varphi$ fluctuations is equivalent to the Schr\"{o}dinger equation of a quantum mechanical model with potential
\begin{eqnarray}
V(\zeta) & = & -\frac{1}{4} \frac{\zeta^2+6}{(\zeta^2+1)^2} - \lambda \frac{1}{\zeta^2+1}, \label{Schrodinger2}
\end{eqnarray}
where in the body of the paper the supersymmetry breaking potential is zero, $\lambda=0$.
The first term can be written in terms of a superpotential as
\begin{eqnarray}
V(\zeta) & = & W^2(\zeta)+W'(\zeta), \nonumber \\
W(\zeta) & = & \frac{1}{\zeta} -\frac{1}{2} \frac{\zeta}{1+\zeta^2}.
\end{eqnarray}
Since the superpotential is  singular, supersymmetry does not guarantee \cite{Panigrahi:1993zy} that the spectrum is positive-semidefinite when $\lambda$ vanishes. In fact, the spectrum contains a low lying, narrow ground state that continues to exist when $\lambda = 0$. For $\zeta>>1$ the potential takes the conformal quantum mechanics form Eq.(\ref{conformal}) with $\alpha=-1/4-\lambda$. For positive values of $\lambda$ there exist a tower of an infinite number of negative energy states above the ground state. For small values of $\lambda$, it has been numerically verified for the ground state and small $n$ excited states  that the bound state spectrum  takes the form
\begin{eqnarray}
E_0 & = & -0.56, \nonumber \\
E_n & = & - C_{2}\, e^{-\frac{n \pi}{\sqrt{\lambda}}},
\end{eqnarray}
with the constant $C_2=0.67$, as depicted in Fig.(\ref{fig:eigenvaluesII}). The ground state wave function is displayed in the left panel of Fig.(\ref{fig:cutoffII}) for $\zeta_\Lambda= 10$ and $\lambda=0.1$. The right panel shows the dependence of the ground state energy on $\zeta_\Lambda$ for $\lambda=0.1$.

\begin{figure*}
\begin{center}
$\begin{array}{cc}
\includegraphics[scale=0.7]{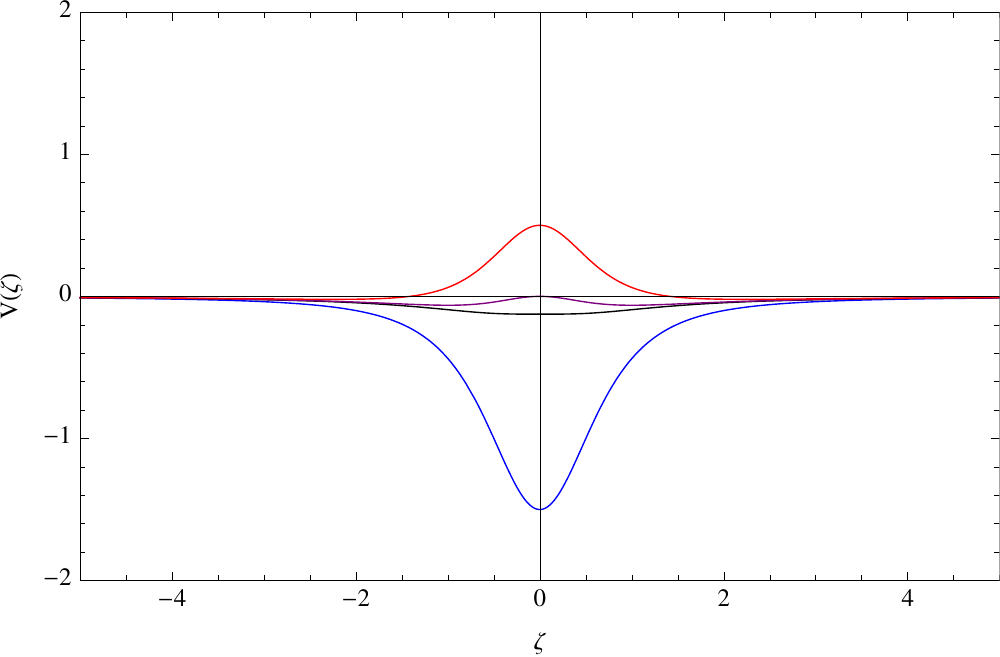} \,\,\,\,\,\,\,&
\includegraphics[scale=0.7]{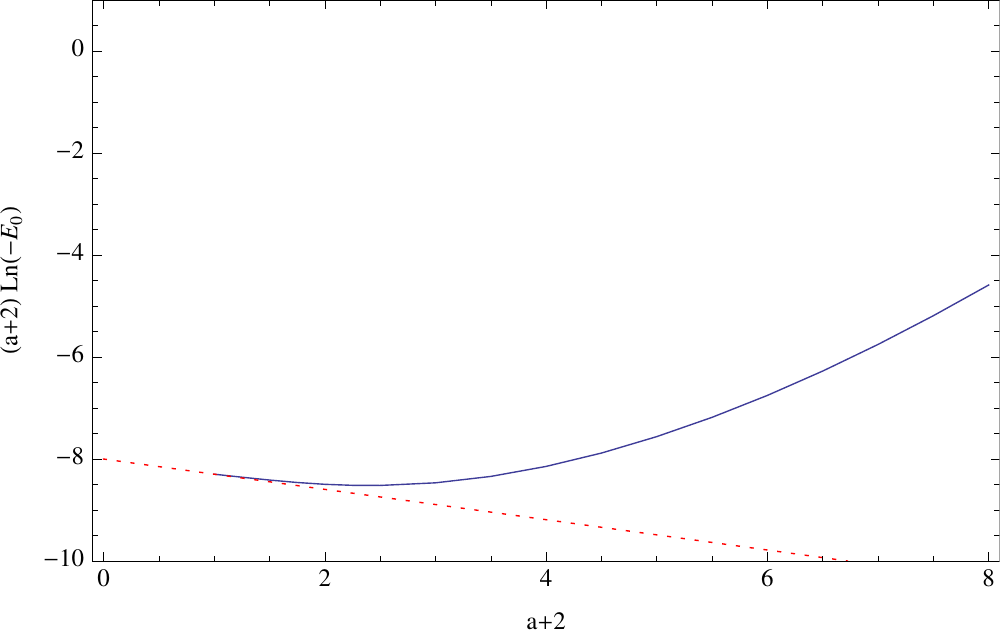}\\
\end{array}$
\caption{Left panel: potential as a function of $\zeta$ for $\lambda=0$ and $a=6$ (blue curve), $a=-1/2$ (black curve), $a=0$ (purple curve), and $a=-2$ (red curve). Right panel: ground state energy as a function of  $(a+2)$ (blue solid curve). Note that $a+2=0$ for Model 1 and $a+2=8$ for Model 2 so that the graph interpolates between the two models. The red dotted curve indicates the non-perturbative asymptotic behavior according to Eq.(\ref{nonperturbeq}) as $a$ approaches the value $-2$ from above. There is no bound state remaining for $a \le -2$.}
\label{fig:transition}
\end{center}
\end{figure*}

Note that the potentials of Models 1 and 2 can be continuously deformed into each other using the interpolating potential
\begin{eqnarray}
V(\zeta) & = & -\frac{1}{4}\frac{\zeta^2+a}{(\zeta^2+1)^2} - \lambda \frac{1}{\zeta^2+1},
\end{eqnarray}
by varying the parameter $a$ from $-2$ to $6$. At some intermediate value of $a$ the spectra of Models 1 and 2 undergo a  rearrangement so that they match up with each other. In particular, while lowering $a$ from $6$ to $-2$ thus flowing from Model 2 to Model 1,  the deep lying ground state of Model 2 shifts up  and widens to become the lowest lying state of the geometric spectrum of Model 1, while the excited states in the geometric sector of the spectrum of Model 2 must widen to match up with the corresponding exited states in Model 1. This is readily apparent for the flow of the ground state as seen in Fig.(\ref{fig:cutoffII}) as it evolves into the ground state as seen in Fig.(\ref{fig:cutoff}).
For $\lambda=0$, the groundstate energy approaches zero exactly as the parameter $a$ approaches the value $-2$ from above according to
\begin{eqnarray}
E_0 & = & - C e^{- \frac{8}{a+2}}, \label{nonperturbeq}
\end{eqnarray}
for values of $a$ near $-2$, as shown in the right panel of Fig.(\ref{fig:transition}). The constant $C$ is numerically determined to have the value $C=0.743$. While Model 1 ($a=-2$) does not exhibit a bound state for $\lambda=0$, a bound state does exist for $a>-2$.

\newpage
\end{document}